\documentclass[%
 reprint,
superscriptaddress,
 amsmath,amssymb,
 aps,
pra,
]{revtex4-1}

\usepackage{graphicx}
\usepackage{dcolumn}
\usepackage{bm}

\begin{document}


\title{Simple model for sequential multiphoton ionization by ultra-intense x-rays}

\author{Xiang Li}
 \email{xiangli@slac.stanford.edu}
 \altaffiliation[Present address:\ ]{Linac Coherent Light Source, SLAC National Accelerator Laboratory, Menlo Park, California 94025, USA.}
 \affiliation{%
 	J.R. Macdonald Laboratory, Department of Physics, Kansas State University, Manhattan, Kansas 66506, USA\\
 }%
 
\author{Rebecca Boll}
\affiliation{%
	European XFEL, Holzkoppel 4, 22869 Schenefeld, Germany\\
}%

\author{Daniel Rolles}

 \affiliation{%
 	J.R. Macdonald Laboratory, Department of Physics, Kansas State University, Manhattan, Kansas 66506, USA\\
 }%
 
 \author{Artem Rudenko}
 
 \affiliation{%
 	J.R. Macdonald Laboratory, Department of Physics, Kansas State University, Manhattan, Kansas 66506, USA\\
 }%

\date{\today}

\begin{abstract}
A simple model for sequential multiphoton ionization by ultra-intense x-rays is presented. The derived scaling of the ion yield with pulse energy quantitatively reproduces the experimental data, which shows that the ion yield increases according to the "power law" behavior typical of multiphoton ionization, followed by saturation at high pulse energies. The calculated average time interval between ionizations for producing ions at a certain charge state is found to be proportional to the pulse duration and independent of all other x-ray pulse parameters. This agrees with previous studies where the kinetic energy of fragment ions with a given charge state produced by intense x-ray ionization of molecules was found to be independent of the pulse energy, but to increase with smaller pulse duration due to the smaller time interval between ionizations.
\end{abstract}

\pacs{Valid PACS appear here}
\maketitle


\section{\label{sec:intro}Introduction}
At photon energies smaller than 100 keV, photoabsorption and scattering are the two major processes underlying the interaction of x-rays with matter, with photoabsorption being several orders of magnitude more probable than scattering at photon energies below 10 keV. In an ultra-intense pulse generated by an x-ray free-electron laser (XFEL), which typically contains $\sim10^{12}$ photons within a pulse duration of a few tens of femtoseconds, multiple x-ray photons can be absorbed by an atom \cite{sorokin_photoelectric_2007,young_femtosecond_2010} or a molecule \cite{hoener_ultraintense_2010}, generating extremely high-charged ionic states \cite{rudek_ultra-efficient_2012,rudenko_femtosecond_2017}. If the x-ray photon energy is high enough, the process usually occurs sequentially, and each of the absorption events leads to ionization, as confirmed by many experimental studies of the interaction of XFEL pulses with atoms \cite{young_femtosecond_2010,rudek_ultra-efficient_2012,fukuzawa_deep_2013} and molecules \cite{hoener_ultraintense_2010,erk_ultrafast_2013,erk_imaging_2014,motomura_charge_2015,nagaya_ultrafast_2016,rudenko_femtosecond_2017}. Such sequential multiphoton ionization is also the source of radiation damage in coherent imaging experiments with XFELs \cite{neutze_potential_2000,barty_ultrafast_2008,nass_indications_2015,galli_electronic_2015}. Theoretically, the interaction of ultra-intense x-rays with atoms and molecules, driven by the sequential multiphoton ionization process, has mostly been treated with ab initio approaches based on “rate-equation” \cite{rohringer_x-ray_2007, young_femtosecond_2010} and Monte Carlo methods \cite{son_monte_2012,ho_theoretical_2014,hao_efficient_2015,inhester_x-ray_2016}, and good agreement with the experimental results has been achieved \cite{young_femtosecond_2010,rudek_ultra-efficient_2012,fukuzawa_deep_2013,rudenko_femtosecond_2017,rudek_relativistic_2018}. 

In this paper, we introduce a simple model to describe the sequential multiphoton ionization process. The scaling of ion yields with pulse energy and the average time interval between ionizations are derived without the need to implement computationally intensive ab initio calculations. By assuming the XFEL pulse to have a Gaussian temporal profile, the number of photoabsorptions to be n, and the average photoabsorption cross section for each ionization step to be $\sigma$, the differential probability for n photoabsorption events to occur at times $t_1$, $t_2$, $\ldots$, $t_n$ is derived. Based on this probability, we obtain the functional dependence of the probability for sequential n-photon ionization to occur, as well as the average timing for each of these n ionization steps. The functional dependence of the sequential n-photon ionization probability is then used to calculate the scaling of the ion yield with pulse energy, which is found to quantitatively describe the behaviour of the measured ion yields showing a "power-law" increase followed by saturation. The average time interval between ionizations for producing a given charge state is found to depend only on the pulse duration with a proportional relation, and not on any other pulse parameters, which is consistent with the conclusion drawn from a previous experiment studying ultra-intense hard x-ray ionization of CH$_3$I molecules \cite{li_pulse_inrev}. In that experiment, the kinetic energy of iodine ion fragments was found to be independent of pulse energy, and to increase with shorter pulse durations because of the decreased time interval between ionizations. In addition to providing general insight on the process of sequential multiphoton ionization, the model can be used to simulate the charge buildup process leading to the explosion of molecules, which is relevant, e.g. for Coulomb explosion imaging at XFEL facilities.

\section{\label{sec:seq}Sequential multiphoton ionization model}

For the simplest case of two-photon ionization, the differential probability for the two photons being absorbed with a time interval $\triangle t$ is derived in Ref. \cite{li_coinc_inrev}. This differential probability is then used to calculate the probability distribution of the kinetic energy release of the ionized molecules. Here we study the general case where the interaction involves the sequential absorption of an arbitrary number of photons. Consider an arbitrary final interaction product requiring a sequence of n photoaborptions from a Gaussian XFEL pulse with photon flux $F(t)\ =\ fe^{-4ln2(\frac{t}{\tau})^{2}}$, where $f$ is the peak photon flux and $\tau$ is the pulse duration (full width at half maximum). And further assume an average photoabsorption cross section $\sigma$ for each ionization step. The differential probability for the sequence of n ionizations to occur at times $t_{1},\ t_{2},\ …,\ t_{n}$ is 
\begin{alignat}{1}
	&\frac{dP}{dt_{1}dt_{2}\cdots dt_{n}}\nonumber\\
	 &= e^{-\int_{-\infty}^{t_{1}}F(t)\sigma dt}F(t_{1})\sigma e^{-\int_{t_{1}}^{t_{2}}F(t)\sigma dt}F(t_{2})\sigma\nonumber\\
	&{\ \ \ }\cdots e^{-\int_{t_{n-1}}^{t_{n}}F(t)\sigma dt}F(t_{n})\sigma e^{-\int_{t_{n}}^{\infty}F(t)\sigma dt}\nonumber\\
	&=\ \sigma^{n}F(t_{1})F(t_{2})\cdots F(t_{n})e^{-\int_{-\infty}^{\infty}F(t)\sigma dt}\nonumber\\
	&=\ e^{-\sqrt{\frac{\pi}{4ln2}}\sigma \tau f}\sigma^{n}F(t_{1})F(t_{2})\cdots F(t_{n}).
\end{alignat}

The differential probability for the sequence of n ionizations to occur at time intervals $\triangle t_{2,1},\ \triangle t_{3,1},\ \ldots,\ \triangle t_{n,1}$ (with $\triangle t_{n,1}\ =\ (t_{n}\ - t_{1}),\ i\ =\ 2,\ 3,\ \ldots,\ n$) is
\begin{alignat}{1}\label{seq_diff_prob}
	&\frac{dP}{d\triangle t_{2,1}d\triangle t_{3,1}\cdots d\triangle t_{n,1}}\nonumber\\
	 &=\ \int_{-\infty}^{\infty}\int_{t_{1}}^{\infty}\cdots \int_{t_{n-1}}^{\infty}\frac{dP}{dt_{1}dt_{2}\cdots dt_{n}}\delta[\triangle t_{2,1}\ -\ (t_{2}\ - t_{1})]\nonumber\\
	&{\ \ \ \ \ }\delta[\triangle t_{3,2}\ -\ (t_{3}\ - t_{2})]\cdots \delta[\triangle t_{n,n-1}\ -\ (t_{n}\ - t_{n-1})]\nonumber\\
	&{\ \ \ }dt_{n}\cdots dt_{2}dt_{1}\nonumber\\
	&=\ e^{-\sqrt{\frac{\pi}{4ln2}}\sigma \tau f}\sigma^{n}\int_{-\infty}^{\infty}\int_{t_{1}}^{\infty}\cdots \int_{t_{n-1}}^{\infty}F(t_{1})\nonumber\\
	&{\ \ \ \ \ }\delta[\triangle t_{2,1}\ -\ (t_{2}\ - t_{1})]F(t_{2})\delta[\triangle t_{3,2}\ -\ (t_{3}\ - t_{2})]\cdots\nonumber\\
	&{\ \ \ \ \ } F(t_{n-1})\delta[\triangle t_{n,n-1}\ -\ (t_{n}\ - t_{n-1})]F(t_{n})dt_{n}\cdots dt_{2}dt_{1}\nonumber\\
	&=\ e^{-\sqrt{\frac{\pi}{4ln2}}\sigma \tau f}\sigma^{n}f^{n}\int_{-\infty}^{\infty}e^{\big[-\frac{-4ln2}{\tau^{2}}[t_{1}^{2}\ +\ (t_{1}\ +\ \triangle t_{2,1})^{2}\ +\ \cdots\ +}\nonumber\\
	&{\ \ \ \ \ }^{(t_{1}\ +\ (\triangle t_{2,1}\ +\ \triangle t_{3,2}\ +\ \cdots\ +\ \triangle t_{n,n-1}))^{2}]\big]}dt_{1}\nonumber\\
	&=\ e^{-\sqrt{\frac{\pi}{4ln2}}\sigma \tau f}\sigma^{n}f^{n}e^{4ln2\frac{A^{2}\ -\ nB}{n\tau^{2}}},
\end{alignat}
where $A\ =\ \triangle t_{2,1}\ +\ \triangle t_{3,1}\ +\ \cdots\ +\ \triangle t_{n,1}$ and $B\ =\ \triangle t_{2,1}^{2}\ +\ \triangle t_{3,1}^{2}\ +\ \cdots\ +\ \triangle t_{n,1}^{2}$, with $\triangle t_{2,1}\ \leq\ \triangle t_{3,1}\ \leq\ \ \cdots\ \leq\ \triangle t_{n,1}$.

\subsection{\label{sub:prob} Probability for sequential n-photon ionization and scaling of the ion yield with pulse energy}

By integrating Eq.~(\ref{seq_diff_prob}) over all possible time intervals, the probability $P_{n}$ to get to a certain interaction product requiring $n$-photon absorption can be obtained:
\begin{align}\label{seq_prob}
	P_{n}\ =\ ae^{-bf}f^{n},
\end{align}
where $a$ is a constant depending on the photon number $n$ and pulse duration $\tau$, and $b\ =\ \sqrt{\frac{\pi}{4ln2}}\sigma\tau$. It is worth noting that Eq. \ref{seq_prob} takes a similar form as Eq. (6) in Ref. \cite{guichard_multiple_2013}, which is derived by analytically solving rate equations.

The equation above can be rewritten in terms of the per-shot ion yield $Y_{n}$ (which is related to $P_{n}$ by $Y_{n}\ =\ NP_{n}$, with $N$ being the number of target atoms or molecules in the interaction volume) and pulse energy $E_{pls}$ (which is the quantity measured in the experiment, and can be approximately related to the photon flux $f$ by $E_{pls}\ = \frac{\tau A E_{pho}}{c_{t}}f$, with $E_{pho}$ being the photon energy, $A$ being the effective x-ray focal area, and  $c_{t}$ the beamline transmission coefficient):
\begin{align}\label{seq_y}
	Y_{n}\ =\ ce^{-dE_{pls}}E_{pls}^{n},
\end{align}
where $c$ is the constant depending on $n$, $\tau$, $N$, $c_{t}$ and $A$, and $d\ =\ \frac{\sqrt{\frac{\pi}{4ln2}}\sigma c_{t}}{AE_{pho}}$.

\subsection{\label{sub:rate} Average time interval of the n sequential ionizations}
By normalizing $\frac{dP}{d\triangle t_{2,1}d\triangle t_{3,1}\cdots d\triangle t_{n,1}}$ in Eq.~(\ref{seq_diff_prob}) such that 
\begin{align}
	&\int_{0}^{\infty}\int_{0}^{\triangle t_{n,1}}\cdots\int_{0}^{\triangle t_{3,1}} G_{n}\frac{dP}{d\triangle t_{2,1}\ d\triangle t_{3,1}\cdots d\triangle t_{n,1}}\nonumber\\&{\ \ \ }d\triangle t_{2,1}\cdots d\triangle t_{n-1,1}d\triangle t_{n,1}\nonumber\\
	&=\ \int_{0}^{\infty}\int_{0}^{\triangle t_{n,1}}\cdots\int_{0}^{\triangle t_{3,1}} C_{n}e^{4ln2\frac{A^{2}\ -\ nB}{n\tau^{2}}}\nonumber\\&{\ \ \ } d\triangle t_{2,1}\cdots d\triangle t_{n-1,1}d\triangle t_{n,1}\nonumber\\
	&=\ 1,
\end{align}
the probability distribution $D_{n}(\triangle t_{2,1},\ \triangle t_{3,1},\ \cdots,\ \triangle t_{n,1})$ of the time interval sequence $\triangle t_{2,1}$, $\triangle t_{3,1}$, $\cdots$, $\triangle t_{n,1}$ can be obtained as
\begin{align}\label{prob_dist_dts}
	D_{n}(\triangle t_{2,1},\ \triangle t_{3,1},\ \cdots,\ \triangle t_{n,1})\ =\ C_{n}e^{4ln2\frac{A^{2}\ -\ nB}{n\tau^{2}}},
\end{align}
where $C_{n}$ is the normalization constant. $C_{n}$ does not depend on the peak intensity f and the cross section $\sigma$ because they are only in the constant $e^{-\sqrt{\frac{\pi}{4ln2}}\sigma \tau f}\sigma^{n}f^{n}$ of the unnormalized Eq.~(\ref{seq_diff_prob}) and are canceled out with the normalization procedure. So the probability distribution $D_{n}(\triangle t_{2,1},\ \triangle t_{3,1},\ \cdots,\ \triangle t_{n,1})$ in Eq.~(\ref{prob_dist_dts}) only depends on the pulse duration $\tau$, which makes the average time interval between ionizations to be derived from this probability distribution only depends on the pulse duration as well. The sequential n-photon ionization therefore becomes more likely with larger peak photon flux as shown by Eq.~(\ref{seq_prob}), but the average time interval between ionizations stays the same as long as the pulse duration is not changed.

In the following, Eq.~(\ref{prob_dist_dts}) will be applied to the specific cases of sequential two- and three-photon ionizations. For sequential two-photon ionizations, the probability distribution for the time interval $\triangle t_{2,1}$ is 
\begin{align}\label{prob_dist_dts2}
	D_{2}(\triangle t_{2,1})\ =\ C_{2}e^{-2ln2\frac{\triangle t_{2,1}^{2}}{\tau^{2}}},
\end{align}
with the normalization constant $C_{2} \ = \frac{2\sqrt{2ln2}}{\sqrt{\pi}\tau}$. From Eq.~(\ref{prob_dist_dts2}), the average time interval between the two photoionizations is 
\begin{align}
	\overline{\triangle t_{2,1}}\ &=\ \int_{0}^{\infty}\triangle t_{2,1}D_{2}(\triangle t_{2,1})\ d\triangle t_{2,1}\nonumber\\
	&=\ \frac{\tau}{\sqrt{2\pi ln2}},
\end{align}. 
For example, if a certain interaction product is created by two photoionizations within a 30 fs XFEL pulse, the average time interval is $\overline{\triangle t_{2,1}}\ =\ \frac{1}{\sqrt{2\pi ln2}}\ \times\ 30\ fs \approx\ 14\ fs$.

For three sequential photoionizations, the probability distribution for the time intervals $\triangle t_{2,1}$ and $\triangle t_{3,1}$ is 
\begin{align}\label{prob_dist_dts3}
	D_{3}(\triangle t_{2,1},\ \triangle t_{3,1})\ =\ C_{3}e^{-4ln2\frac{-2\triangle t_{2,1}^{2}\ +\ 2\triangle t_{2,1}\triangle t_{3,1}\ -\ 2\triangle t_{3,1}^{2}}{3\tau^{2}}},
\end{align}
with the normalization constant $C_{3} \ = \frac{48ln2}{\sqrt{3}\pi\tau^{2}}$. From Eq.~(\ref{prob_dist_dts3}), the average time interval between the first and last photoionizations can be calculated: 
\begin{align}
	\overline{\triangle t_{3,1}}\ &=\ \int_{0}^{\infty}\int_{0}^{\triangle t_{3,1}}\triangle t_{3,1}D_{3}(\triangle t_{2,1},\ \triangle t_{3,1})d\triangle t_{2,1}d\triangle t_{3,1}\nonumber\\
	&=\ \frac{3}{2\sqrt{2\pi ln2}}\tau,
\end{align}. 
If a certain interaction product is created by three photoionizations within a 30 fs XFEL pulse, the average time interval is thus $\overline{\triangle t_{3,1}}\ =\ \frac{3}{2\sqrt{2\pi ln2}}\ \times\ 30 fs\ \approx\ 22\ fs$.

The average time intervals for interactions with larger number of sequential photoionizations can be calculated analogously as for the sequential three-photon ionization. For the general case of sequential n-photon ionization, it is shown in the following that the average time interval is proportional to the pulse duration $\tau$. 

Introducing the scaled time interval $\triangle t'_{i,1}\ =\ \frac{\triangle t_{i,1}}{\tau}$, with $i = 2,\ 3,\ \cdots,\ n$, the distribution for these scaled time intervals is
\begin{align}\label{prob_dist_dts'}
	D'_{n}(\triangle t'_{2,1},\ \triangle t'_{3,1},\ \cdots,\ \triangle t'_{n,1})\ =\ C'_{n}e^{4ln2\frac{A'^{2}\ -\ nB'}{n}},
\end{align}
where $C'_{n}$ is the normalization constant, which does not depend on $\tau$, $A'\ =\ \triangle t'_{2,1}\ +\ \triangle t'_{3,1}\ +\ \cdots\ +\ \triangle t'_{n,1}$ and $B'\ =\ \triangle t_{2,1}^{'2}\ +\ \triangle t_{3,1}^{'2}\ +\ \cdots\ +\ \triangle t_{n,1}^{'2}$, with $\triangle t'_{2,1}\ \leq\ \triangle t'_{3,1}\ \leq\ \ \cdots\ \leq\ \triangle t'_{n,1}$. Since $D'_{n}(\triangle t'_{2,1},\ \triangle t'_{3,1},\ \cdots,\ \triangle t'_{n,1})$ does not depend on $\tau$, neither does the average time interval $\overline{\triangle t'_{n,1}}$, which can be obtained from 
\begin{align}
	\overline{\triangle t'_{n,1}}\ &=\ \int_{0}^{\infty}\int_{0}^{\triangle t'_{n,1}}\cdots\int_{0}^{\triangle t'_{3,1}}\triangle t'_{n,1}D'_{n}(\triangle t'_{2,1},\ \triangle t'_{3,1},\nonumber\\&{\ \ \ } \cdots,\ \triangle t'_{n,1}) d\triangle t'_{2,1}\cdots d\triangle t'_{n-1,1}d\triangle t'_{n,1}.
\end{align}
The unscaled average time interval is then $\overline{\triangle t_{n,1}}\ =\ \overline{\triangle t'_{i,1}}\tau\ \propto \tau$.

\section{\label{sec:comp} Application of the simple model}

\subsection{\label{sub:prob_c} Scaling of the ion yield with pulse energy}
\begin{figure}[htb]
	
	\includegraphics[width=3.3in,height=2.2in]{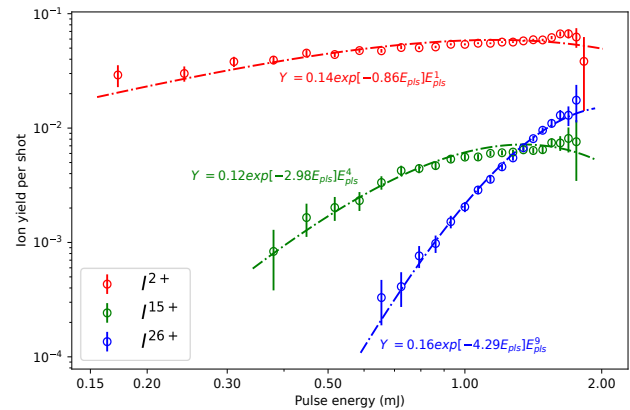}
	
	\caption{Yield of $I^{2+}$, $I^{15+}$, and $I^{26+}$ as a function of pulse energy, produced by the interaction between CH$_{3}$I molecules and 2 keV x-rays generated by the European X-ray Free-Electron Laser. The pulse energy was measured upstream of the the Small Quantum Systems instrument, in which the experiment was performed. These data are "non-coincident", i.e., integrated over all carbon ion charge states. The dashed lines are fits using Eq.~(\ref{seq_y}), with the fitted functions displayed in the figure.}
	
	\label{yield_plseng}
\end{figure}

Eq.~(\ref{seq_y}) can be used to describe the dependence of the ion yield on the x-ray pulse energy. In Fig.~\ref{yield_plseng}, the experimental ion yield per-shot for three representative ions, $I^{2+}$, $I^{15+}$, and $I^{26+}$, created by the interaction between CH$_{3}$I molecules and 2 keV x-rays generated by the European X-ray Free-Electron Laser, are plotted as a function of pulse energy as measured upstream of the Small Quantum Systems (SQS) instrument, together with fits of the function in Eq.~(\ref{seq_y}), which is derived from our simple model.

The parameters c and d, which are dependent on experimental settings (such as the beam focal area) are used as fitting parameters since they can not be easily pre-determined. n, which is the average number of photoabsorptions, can be pre-estimated by energy arguments. One photoabsorption is enough to reach $I^{2+}$, thus $n$ was set to 1 to fit the $I^{2+}$ ion yield. The minimum number of photoabsorptions required to reach the other two charge states can be estimated by applying energy conservation. Take the production of $I^{26+}$ as an example. The total energy of electrons in the neutral CH$_{3}$I molecule is approximately the sum of the energy of the electrons in the individual atoms, which is -198.369 keV. When $I^{26+}$ is produced, its coincident carbon partner most likely has 4 charges, and all three hydrogen fragments are also charged, as shown in Fig.~5.1 and 6.2 of Ref. \cite{li_molecular_2019}. The total energy of electrons in the fragments [$I^{26+}$, $C^{4+}$, $3\ \times\ H^{+}$] is -187.646 keV. The average energy of photoelectrons ionized for the production of $I^{26+}$ is 0.599 keV. (These energy calculations were made with the Hartree-Fock method through Cowan's atomic code \cite{cowan_code, cowan_book_1981}.) From energy conservation, the total energy of absorbed photons subtracted by the total energy of photoelectrons, which is $(n\ \times 2\ -\ n\ \times\ 0.599)$ keV, should be equal to or larger (to account for radiative decay processes) than the energy change in the molecular system, which is $(198.369\ -\ 187.646)$ keV. The minimum of n satisfying this requirement is 8. $n\ =\ 9$ is used to account for possible radiative decays and the fact that the total energy of the neutral molecule is smaller than that estimated with the independent-atom model. (A more precise estimate of the number of photons needed to reach a particular charge state requires a detailed knowledge of the average decay pathways involved.) With n thus determined to be 1, 4 and 9 for $I^{2+}$, $I^{15+}$ and $I^{26+}$, respectively, Eq.~(\ref{seq_y}) is used to fit the experimental data. The resulting function for each charge state is displayed in Fig.~\ref{yield_plseng}. The fitting values for $d$ contain information about the x-ray transmission, focal area and photoabsorption cross section. For x-rays with a photon energy of 2 keV, the photoabsorption cross section $\sigma$ for the CH$_{3}$I molecule is approximately that of iodine, which is 0.4 Mbarn \cite{berger_xcom_1998}. Using this value for $\sigma$, a typical beamline transmission coefficient $c_{t}$ of 0.8, and the fitting values of $d$ for $I^{2+}$, $I^{15+}$ and $I^{26+}$, the calculated effective focal areas are about 124 $\mu m^{2}$, 36 $\mu m^{2}$ and 25 $\mu m^{2}$, respectively. These values are larger than the expected focal area $\sim 1\ \mu m^{2}$, because with the applied fomula $E_{pls}\ = \frac{\tau A E_{pho}}{c_{t}}f$, the effective focal area $A$ is the average beam area in the whole interaction region, which is much larger than the focal area at the beam center. The observation that higher-charged ions have smaller effective focal areas can be attributed to the fact that such ions require higher x-ray intensity and hence are produced in a more confined region around the x-ray focus. If the beamline transmission and effective focal area are known, the fitting value for $d$ can also be used to calculate the (average) photoabsorption cross section. It is worth noting that such calculations of the effective focal area $A$ and average cross section assume that the average number n of photoabsorptions is known beforehand. For more precise calculations, a more accurate method to estimate n is required. In addition, if the parameter d can be determined by independent measurements, the number of photoabsorptions n and the parameter c can be used as fitting parameters, providing a way to measure the average number of photoabsorptions required to reach a certain interaction product.

According to Eq.~(\ref{seq_y}), the yield of a particular charge state requiring $n$ photoabsorptions initially follows a power law, i.e., $Y_{n}\ \propto\ E_{pls}^{n}$, in the low pulse energy range. At higher pulse energies, the exponential factor $e^{-dE_{pls}}$ becomes non-negligible, and the ion yield starts to saturate and eventually decreases with increasing pulse energy. This describes the fact that if the pulse energy continues to increase, only the highest energetically achievable charge state, which can not be further ionized, is left as the final interaction product, while all other ion charge states are depleted. 

\subsection{\label{sub:rate_c} Average time interval of the n sequential ionizations}
Since the probability for a single photoionization increases with the photon flux or pulse intensity, it is tempting to conclude that the average time interval of the n sequential ionizations decreases with theses two quantities. However, within the sequential multiphoton ionization model introduced in subsection \ref{sub:rate}, the time interval is proportional to the pulse duration $\tau$ and independent of all other pulse parameters such as photon flux.

The average time interval between ionizations can not be easily studied with experiments using atomic targets. For molecular targets, however, it is reflected in the fragment ion kinetic energies. If the time interval is small, the ion kinetic energy would be large because with ionizations separated by small time intervals, a certain ion charge state is reached at a short internuclear distance, which causes stronger Coulomb repulsion. The dependence of the average time interval between ionizations on the x-ray pulse parameters can therefore be studied by measuring the kinetic energy of ions produced from ionization of molecules by x-rays with different pulse energies or durations. In a previous study \cite{li_pulse_inrev}, where we studied the pulse energy and pulse duration dependence of ultra-intense hard x-ray ionization of CH$_3$I molecules, the iodine ion kinetic energy was found to be independent of pulse energy, but increases with shorter pulse duration. This experimental observation supports the conclusion reached from the sequential multiphoton ionization model that the average time interval between ionizations is proportional to the pulse duration $\tau$ and independent of all other pulse parameters. An intuitive picture which explains such pulse parameter dependence can be found in \cite{li_pulse_inrev}.

As an example of how the sequential model can be used in practice, we briefly discuss in the following how it can be applied to facilitate the interpretation of Coulomb explosion imaging experiments at XFEL facilities. To implement the Coulomb explosion imaging technique, Coulomb explosion simulations are required to compare with experimental observations. If the molecule is assumed to be instantaneously charged up, there will be a mismatch between the simulation and experimental observation due to the neglected charge buildup and redistribution processes \cite{li_molecular_2019,li_coulomb_tosub}. To make a more accurate comparison possible, these two processes need to be included and the sequential multiphoton ionization model can be used to help simulate the charge buildup process. In particular, the experimental parameters such as the average number of photons required to reach a certain fragmentation product can be estimated by fitting the experimental data with Eq. (\ref{seq_y}), and the corresponding time interval between sequential photoionizations can be calculated according to the discussion in subsection \ref{sub:rate}. These numbers can then be incorporated for simulating the charge buildup process.

\section{\label{sec:sum} Summary}
A simple sequential multiphoton ionization model is presented, from which the probability for sequential n-photon ionization, the scaling of the ion yield with pulse energy, as well as the average time interval between ionizations can be calculated. The derived scaling of the ion yield with pulse energy in Eq.~(\ref{seq_y}) contains a term describing the increase of the ion yield with pulse energy, $E_{pls}$ according to the power law $E_{pls}^n$, as expected from a typical n-photon-absorption nonlinear process, and an exponentially decaying factor that describes saturation and depletion. The exponential factor can be neglected at small pulse energies but dominates the behavior at large pulse energies. Eq.~(\ref{seq_y}) is shown to quantitatively describe the "power-law"-type increase followed by saturation observed in the experimental data. The average time interval between ionizations calculated within the model is found to be proportional to the pulse duration and independent of all other pulse parameters. This is consistent with our previous experimental observation that the kinetic energy of fragment ions produced by the ionization of molecules with intense XFEL pulses is independent of pulse energy but increases with smaller pulse duration due to the smaller time interval between ionizations.

\section{Acknowledgements}
This work was funded by the U.S.~Department of Energy, Office of Science, Basic Energy Sciences (BES), Division of Chemical Sciences, Geosciences, and Biosciences, under Grants No.~DE-FG02-86ER13491 (D.R. and A.R.) and DE-SC0019451 (X.L.). We thank the colleagues from the European XFEL and University of Frankfurt, who participated in the beamtime at the European XFEL, during which the experimental data shown in Fig.~\ref{yield_plseng} was collected, which is used to demonstrate the application of our model. We also thank the scientific and technical staff of the European XFEL for their help and hospitality during the beamtime. We acknowledge valuable discussions with Ludger Inhester, Sang-Kil Son and Robin Santra from the Center for Free-Electron Laser Science at DESY.


\begin{thebibliography}{28}%
	\makeatletter
	\providecommand \@ifxundefined [1]{%
		\@ifx{#1\undefined}
	}%
	\providecommand \@ifnum [1]{%
		\ifnum #1\expandafter \@firstoftwo
		\else \expandafter \@secondoftwo
		\fi
	}%
	\providecommand \@ifx [1]{%
		\ifx #1\expandafter \@firstoftwo
		\else \expandafter \@secondoftwo
		\fi
	}%
	\providecommand \natexlab [1]{#1}%
	\providecommand \enquote  [1]{``#1''}%
	\providecommand \bibnamefont  [1]{#1}%
	\providecommand \bibfnamefont [1]{#1}%
	\providecommand \citenamefont [1]{#1}%
	\providecommand \href@noop [0]{\@secondoftwo}%
	\providecommand \href [0]{\begingroup \@sanitize@url \@href}%
	\providecommand \@href[1]{\@@startlink{#1}\@@href}%
	\providecommand \@@href[1]{\endgroup#1\@@endlink}%
	\providecommand \@sanitize@url [0]{\catcode `\\12\catcode `\$12\catcode
		`\&12\catcode `\#12\catcode `\^12\catcode `\_12\catcode `\%12\relax}%
	\providecommand \@@startlink[1]{}%
	\providecommand \@@endlink[0]{}%
	\providecommand \url  [0]{\begingroup\@sanitize@url \@url }%
	\providecommand \@url [1]{\endgroup\@href {#1}{\urlprefix }}%
	\providecommand \urlprefix  [0]{URL }%
	\providecommand \Eprint [0]{\href }%
	\providecommand \doibase [0]{http://dx.doi.org/}%
	\providecommand \selectlanguage [0]{\@gobble}%
	\providecommand \bibinfo  [0]{\@secondoftwo}%
	\providecommand \bibfield  [0]{\@secondoftwo}%
	\providecommand \translation [1]{[#1]}%
	\providecommand \BibitemOpen [0]{}%
	\providecommand \bibitemStop [0]{}%
	\providecommand \bibitemNoStop [0]{.\EOS\space}%
	\providecommand \EOS [0]{\spacefactor3000\relax}%
	\providecommand \BibitemShut  [1]{\csname bibitem#1\endcsname}%
	\let\auto@bib@innerbib\@empty
	\bibitem [{\citenamefont {Sorokin}\ \emph {et~al.}(2007)\citenamefont
		{Sorokin}, \citenamefont {Bobashev}, \citenamefont {Feigl}, \citenamefont
		{Tiedtke}, \citenamefont {Wabnitz},\ and\ \citenamefont
		{Richter}}]{sorokin_photoelectric_2007}%
	\BibitemOpen
	\bibfield  {author} {\bibinfo {author} {\bibfnamefont {A.~A.}\ \bibnamefont
			{Sorokin}}, \bibinfo {author} {\bibfnamefont {S.~V.}\ \bibnamefont
			{Bobashev}}, \bibinfo {author} {\bibfnamefont {T.}~\bibnamefont {Feigl}},
		\bibinfo {author} {\bibfnamefont {K.}~\bibnamefont {Tiedtke}}, \bibinfo
		{author} {\bibfnamefont {H.}~\bibnamefont {Wabnitz}}, \ and\ \bibinfo
		{author} {\bibfnamefont {M.}~\bibnamefont {Richter}},\ }\href
	{https://link.aps.org/doi/10.1103/PhysRevLett.99.213002} {\bibfield
		{journal} {\bibinfo  {journal} {Physical Review Letters}\ }\textbf {\bibinfo
			{volume} {99}},\ \bibinfo {pages} {213002} (\bibinfo {year}
		{2007})}\BibitemShut {NoStop}%
	\bibitem [{\citenamefont {Young}\ \emph {et~al.}(2010)\citenamefont {Young},
		\citenamefont {Kanter}, \citenamefont {Krässig}, \citenamefont {Li},
		\citenamefont {March}, \citenamefont {Pratt}, \citenamefont {Santra},
		\citenamefont {Southworth}, \citenamefont {Rohringer}, \citenamefont
		{DiMauro}, \citenamefont {Doumy}, \citenamefont {Roedig}, \citenamefont
		{Berrah}, \citenamefont {Fang}, \citenamefont {Hoener}, \citenamefont
		{Bucksbaum}, \citenamefont {Cryan}, \citenamefont {Ghimire}, \citenamefont
		{Glownia}, \citenamefont {Reis}, \citenamefont {Bozek}, \citenamefont
		{Bostedt},\ and\ \citenamefont {Messerschmidt}}]{young_femtosecond_2010}%
	\BibitemOpen
	\bibfield  {author} {\bibinfo {author} {\bibfnamefont {L.}~\bibnamefont
			{Young}}, \bibinfo {author} {\bibfnamefont {E.~P.}\ \bibnamefont {Kanter}},
		\bibinfo {author} {\bibfnamefont {B.}~\bibnamefont {Krässig}}, \bibinfo
		{author} {\bibfnamefont {Y.}~\bibnamefont {Li}}, \bibinfo {author}
		{\bibfnamefont {A.~M.}\ \bibnamefont {March}}, \bibinfo {author}
		{\bibfnamefont {S.~T.}\ \bibnamefont {Pratt}}, \bibinfo {author}
		{\bibfnamefont {R.}~\bibnamefont {Santra}}, \bibinfo {author} {\bibfnamefont
			{S.~H.}\ \bibnamefont {Southworth}}, \bibinfo {author} {\bibfnamefont
			{N.}~\bibnamefont {Rohringer}}, \bibinfo {author} {\bibfnamefont {L.~F.}\
			\bibnamefont {DiMauro}}, \bibinfo {author} {\bibfnamefont {G.}~\bibnamefont
			{Doumy}}, \bibinfo {author} {\bibfnamefont {C.~A.}\ \bibnamefont {Roedig}},
		\bibinfo {author} {\bibfnamefont {N.}~\bibnamefont {Berrah}}, \bibinfo
		{author} {\bibfnamefont {L.}~\bibnamefont {Fang}}, \bibinfo {author}
		{\bibfnamefont {M.}~\bibnamefont {Hoener}}, \bibinfo {author} {\bibfnamefont
			{P.~H.}\ \bibnamefont {Bucksbaum}}, \bibinfo {author} {\bibfnamefont {J.~P.}\
			\bibnamefont {Cryan}}, \bibinfo {author} {\bibfnamefont {S.}~\bibnamefont
			{Ghimire}}, \bibinfo {author} {\bibfnamefont {J.~M.}\ \bibnamefont
			{Glownia}}, \bibinfo {author} {\bibfnamefont {D.~A.}\ \bibnamefont {Reis}},
		\bibinfo {author} {\bibfnamefont {J.~D.}\ \bibnamefont {Bozek}}, \bibinfo
		{author} {\bibfnamefont {C.}~\bibnamefont {Bostedt}}, \ and\ \bibinfo
		{author} {\bibfnamefont {M.}~\bibnamefont {Messerschmidt}},\ }\href
	{https://www.nature.com/articles/nature09177} {\bibfield  {journal} {\bibinfo
			{journal} {Nature}\ }\textbf {\bibinfo {volume} {466}},\ \bibinfo {pages}
		{56} (\bibinfo {year} {2010})}\BibitemShut {NoStop}%
	\bibitem [{\citenamefont {Hoener}\ \emph {et~al.}(2010)\citenamefont {Hoener},
		\citenamefont {Fang}, \citenamefont {Kornilov}, \citenamefont {Gessner},
		\citenamefont {Pratt}, \citenamefont {Gühr}, \citenamefont {Kanter},
		\citenamefont {Blaga}, \citenamefont {Bostedt}, \citenamefont {Bozek},
		\citenamefont {Bucksbaum}, \citenamefont {Buth}, \citenamefont {Chen},
		\citenamefont {Coffee}, \citenamefont {Cryan}, \citenamefont {DiMauro},
		\citenamefont {Glownia}, \citenamefont {Hosler}, \citenamefont {Kukk},
		\citenamefont {Leone}, \citenamefont {McFarland}, \citenamefont
		{Messerschmidt}, \citenamefont {Murphy}, \citenamefont {Petrovic},
		\citenamefont {Rolles},\ and\ \citenamefont
		{Berrah}}]{hoener_ultraintense_2010}%
	\BibitemOpen
	\bibfield  {author} {\bibinfo {author} {\bibfnamefont {M.}~\bibnamefont
			{Hoener}}, \bibinfo {author} {\bibfnamefont {L.}~\bibnamefont {Fang}},
		\bibinfo {author} {\bibfnamefont {O.}~\bibnamefont {Kornilov}}, \bibinfo
		{author} {\bibfnamefont {O.}~\bibnamefont {Gessner}}, \bibinfo {author}
		{\bibfnamefont {S.~T.}\ \bibnamefont {Pratt}}, \bibinfo {author}
		{\bibfnamefont {M.}~\bibnamefont {Gühr}}, \bibinfo {author} {\bibfnamefont
			{E.~P.}\ \bibnamefont {Kanter}}, \bibinfo {author} {\bibfnamefont
			{C.}~\bibnamefont {Blaga}}, \bibinfo {author} {\bibfnamefont
			{C.}~\bibnamefont {Bostedt}}, \bibinfo {author} {\bibfnamefont {J.~D.}\
			\bibnamefont {Bozek}}, \bibinfo {author} {\bibfnamefont {P.~H.}\ \bibnamefont
			{Bucksbaum}}, \bibinfo {author} {\bibfnamefont {C.}~\bibnamefont {Buth}},
		\bibinfo {author} {\bibfnamefont {M.}~\bibnamefont {Chen}}, \bibinfo {author}
		{\bibfnamefont {R.}~\bibnamefont {Coffee}}, \bibinfo {author} {\bibfnamefont
			{J.}~\bibnamefont {Cryan}}, \bibinfo {author} {\bibfnamefont
			{L.}~\bibnamefont {DiMauro}}, \bibinfo {author} {\bibfnamefont
			{M.}~\bibnamefont {Glownia}}, \bibinfo {author} {\bibfnamefont
			{E.}~\bibnamefont {Hosler}}, \bibinfo {author} {\bibfnamefont
			{E.}~\bibnamefont {Kukk}}, \bibinfo {author} {\bibfnamefont {S.~R.}\
			\bibnamefont {Leone}}, \bibinfo {author} {\bibfnamefont {B.}~\bibnamefont
			{McFarland}}, \bibinfo {author} {\bibfnamefont {M.}~\bibnamefont
			{Messerschmidt}}, \bibinfo {author} {\bibfnamefont {B.}~\bibnamefont
			{Murphy}}, \bibinfo {author} {\bibfnamefont {V.}~\bibnamefont {Petrovic}},
		\bibinfo {author} {\bibfnamefont {D.}~\bibnamefont {Rolles}}, \ and\ \bibinfo
		{author} {\bibfnamefont {N.}~\bibnamefont {Berrah}},\ }\href
	{https://link.aps.org/doi/10.1103/PhysRevLett.104.253002} {\bibfield
		{journal} {\bibinfo  {journal} {Physical Review Letters}\ }\textbf {\bibinfo
			{volume} {104}},\ \bibinfo {pages} {253002} (\bibinfo {year}
		{2010})}\BibitemShut {NoStop}%
	\bibitem [{\citenamefont {Rudek}\ \emph {et~al.}(2012)\citenamefont {Rudek},
		\citenamefont {Son}, \citenamefont {Foucar}, \citenamefont {Epp},
		\citenamefont {Erk}, \citenamefont {Hartmann}, \citenamefont {Adolph},
		\citenamefont {Andritschke}, \citenamefont {Aquila}, \citenamefont {Berrah},
		\citenamefont {Bostedt}, \citenamefont {Bozek}, \citenamefont {Coppola},
		\citenamefont {Filsinger}, \citenamefont {Gorke}, \citenamefont {Gorkhover},
		\citenamefont {Graafsma}, \citenamefont {Gumprecht}, \citenamefont
		{Hartmann}, \citenamefont {Hauser}, \citenamefont {Herrmann}, \citenamefont
		{Hirsemann}, \citenamefont {Holl}, \citenamefont {Hömke}, \citenamefont
		{Journel}, \citenamefont {Kaiser}, \citenamefont {Kimmel}, \citenamefont
		{Krasniqi}, \citenamefont {Kühnel}, \citenamefont {Matysek}, \citenamefont
		{Messerschmidt}, \citenamefont {Miesner}, \citenamefont {Möller},
		\citenamefont {Moshammer}, \citenamefont {Nagaya}, \citenamefont {Nilsson},
		\citenamefont {Potdevin}, \citenamefont {Pietschner}, \citenamefont {Reich},
		\citenamefont {Rupp}, \citenamefont {Schaller}, \citenamefont {Schlichting},
		\citenamefont {Schmidt}, \citenamefont {Schopper}, \citenamefont {Schorb},
		\citenamefont {Schröter}, \citenamefont {Schulz}, \citenamefont {Simon},
		\citenamefont {Soltau}, \citenamefont {Strüder}, \citenamefont {Ueda},
		\citenamefont {Weidenspointner}, \citenamefont {Santra}, \citenamefont
		{Ullrich}, \citenamefont {Rudenko},\ and\ \citenamefont
		{Rolles}}]{rudek_ultra-efficient_2012}%
	\BibitemOpen
	\bibfield  {author} {\bibinfo {author} {\bibfnamefont {B.}~\bibnamefont
			{Rudek}}, \bibinfo {author} {\bibfnamefont {S.-K.}\ \bibnamefont {Son}},
		\bibinfo {author} {\bibfnamefont {L.}~\bibnamefont {Foucar}}, \bibinfo
		{author} {\bibfnamefont {S.~W.}\ \bibnamefont {Epp}}, \bibinfo {author}
		{\bibfnamefont {B.}~\bibnamefont {Erk}}, \bibinfo {author} {\bibfnamefont
			{R.}~\bibnamefont {Hartmann}}, \bibinfo {author} {\bibfnamefont
			{M.}~\bibnamefont {Adolph}}, \bibinfo {author} {\bibfnamefont
			{R.}~\bibnamefont {Andritschke}}, \bibinfo {author} {\bibfnamefont
			{A.}~\bibnamefont {Aquila}}, \bibinfo {author} {\bibfnamefont
			{N.}~\bibnamefont {Berrah}}, \bibinfo {author} {\bibfnamefont
			{C.}~\bibnamefont {Bostedt}}, \bibinfo {author} {\bibfnamefont
			{J.}~\bibnamefont {Bozek}}, \bibinfo {author} {\bibfnamefont
			{N.}~\bibnamefont {Coppola}}, \bibinfo {author} {\bibfnamefont
			{F.}~\bibnamefont {Filsinger}}, \bibinfo {author} {\bibfnamefont
			{H.}~\bibnamefont {Gorke}}, \bibinfo {author} {\bibfnamefont
			{T.}~\bibnamefont {Gorkhover}}, \bibinfo {author} {\bibfnamefont
			{H.}~\bibnamefont {Graafsma}}, \bibinfo {author} {\bibfnamefont
			{L.}~\bibnamefont {Gumprecht}}, \bibinfo {author} {\bibfnamefont
			{A.}~\bibnamefont {Hartmann}}, \bibinfo {author} {\bibfnamefont
			{G.}~\bibnamefont {Hauser}}, \bibinfo {author} {\bibfnamefont
			{S.}~\bibnamefont {Herrmann}}, \bibinfo {author} {\bibfnamefont
			{H.}~\bibnamefont {Hirsemann}}, \bibinfo {author} {\bibfnamefont
			{P.}~\bibnamefont {Holl}}, \bibinfo {author} {\bibfnamefont {A.}~\bibnamefont
			{Hömke}}, \bibinfo {author} {\bibfnamefont {L.}~\bibnamefont {Journel}},
		\bibinfo {author} {\bibfnamefont {C.}~\bibnamefont {Kaiser}}, \bibinfo
		{author} {\bibfnamefont {N.}~\bibnamefont {Kimmel}}, \bibinfo {author}
		{\bibfnamefont {F.}~\bibnamefont {Krasniqi}}, \bibinfo {author}
		{\bibfnamefont {K.-U.}\ \bibnamefont {Kühnel}}, \bibinfo {author}
		{\bibfnamefont {M.}~\bibnamefont {Matysek}}, \bibinfo {author} {\bibfnamefont
			{M.}~\bibnamefont {Messerschmidt}}, \bibinfo {author} {\bibfnamefont
			{D.}~\bibnamefont {Miesner}}, \bibinfo {author} {\bibfnamefont
			{T.}~\bibnamefont {Möller}}, \bibinfo {author} {\bibfnamefont
			{R.}~\bibnamefont {Moshammer}}, \bibinfo {author} {\bibfnamefont
			{K.}~\bibnamefont {Nagaya}}, \bibinfo {author} {\bibfnamefont
			{B.}~\bibnamefont {Nilsson}}, \bibinfo {author} {\bibfnamefont
			{G.}~\bibnamefont {Potdevin}}, \bibinfo {author} {\bibfnamefont
			{D.}~\bibnamefont {Pietschner}}, \bibinfo {author} {\bibfnamefont
			{C.}~\bibnamefont {Reich}}, \bibinfo {author} {\bibfnamefont
			{D.}~\bibnamefont {Rupp}}, \bibinfo {author} {\bibfnamefont {G.}~\bibnamefont
			{Schaller}}, \bibinfo {author} {\bibfnamefont {I.}~\bibnamefont
			{Schlichting}}, \bibinfo {author} {\bibfnamefont {C.}~\bibnamefont
			{Schmidt}}, \bibinfo {author} {\bibfnamefont {F.}~\bibnamefont {Schopper}},
		\bibinfo {author} {\bibfnamefont {S.}~\bibnamefont {Schorb}}, \bibinfo
		{author} {\bibfnamefont {C.-D.}\ \bibnamefont {Schröter}}, \bibinfo {author}
		{\bibfnamefont {J.}~\bibnamefont {Schulz}}, \bibinfo {author} {\bibfnamefont
			{M.}~\bibnamefont {Simon}}, \bibinfo {author} {\bibfnamefont
			{H.}~\bibnamefont {Soltau}}, \bibinfo {author} {\bibfnamefont
			{L.}~\bibnamefont {Strüder}}, \bibinfo {author} {\bibfnamefont
			{K.}~\bibnamefont {Ueda}}, \bibinfo {author} {\bibfnamefont {G.}~\bibnamefont
			{Weidenspointner}}, \bibinfo {author} {\bibfnamefont {R.}~\bibnamefont
			{Santra}}, \bibinfo {author} {\bibfnamefont {J.}~\bibnamefont {Ullrich}},
		\bibinfo {author} {\bibfnamefont {A.}~\bibnamefont {Rudenko}}, \ and\
		\bibinfo {author} {\bibfnamefont {D.}~\bibnamefont {Rolles}},\ }\href
	{https://www.nature.com/articles/nphoton.2012.261} {\bibfield  {journal}
		{\bibinfo  {journal} {Nature Photonics}\ }\textbf {\bibinfo {volume} {6}},\
		\bibinfo {pages} {858} (\bibinfo {year} {2012})}\BibitemShut {NoStop}%
	\bibitem [{\citenamefont {Rudenko}\ \emph {et~al.}(2017)\citenamefont
		{Rudenko}, \citenamefont {Inhester}, \citenamefont {Hanasaki}, \citenamefont
		{Li}, \citenamefont {Robatjazi}, \citenamefont {Erk}, \citenamefont {Boll},
		\citenamefont {Toyota}, \citenamefont {Hao}, \citenamefont {Vendrell},
		\citenamefont {Bomme}, \citenamefont {Savelyev}, \citenamefont {Rudek},
		\citenamefont {Foucar}, \citenamefont {Southworth}, \citenamefont {Lehmann},
		\citenamefont {Kraessig}, \citenamefont {Marchenko}, \citenamefont {Simon},
		\citenamefont {Ueda}, \citenamefont {Ferguson}, \citenamefont {Bucher},
		\citenamefont {Gorkhover}, \citenamefont {Carron}, \citenamefont
		{Alonso-Mori}, \citenamefont {Koglin}, \citenamefont {Correa}, \citenamefont
		{Williams}, \citenamefont {Boutet}, \citenamefont {Young}, \citenamefont
		{Bostedt}, \citenamefont {Son}, \citenamefont {Santra},\ and\ \citenamefont
		{Rolles}}]{rudenko_femtosecond_2017}%
	\BibitemOpen
	\bibfield  {author} {\bibinfo {author} {\bibfnamefont {A.}~\bibnamefont
			{Rudenko}}, \bibinfo {author} {\bibfnamefont {L.}~\bibnamefont {Inhester}},
		\bibinfo {author} {\bibfnamefont {K.}~\bibnamefont {Hanasaki}}, \bibinfo
		{author} {\bibfnamefont {X.}~\bibnamefont {Li}}, \bibinfo {author}
		{\bibfnamefont {S.~J.}\ \bibnamefont {Robatjazi}}, \bibinfo {author}
		{\bibfnamefont {B.}~\bibnamefont {Erk}}, \bibinfo {author} {\bibfnamefont
			{R.}~\bibnamefont {Boll}}, \bibinfo {author} {\bibfnamefont {K.}~\bibnamefont
			{Toyota}}, \bibinfo {author} {\bibfnamefont {Y.}~\bibnamefont {Hao}},
		\bibinfo {author} {\bibfnamefont {O.}~\bibnamefont {Vendrell}}, \bibinfo
		{author} {\bibfnamefont {C.}~\bibnamefont {Bomme}}, \bibinfo {author}
		{\bibfnamefont {E.}~\bibnamefont {Savelyev}}, \bibinfo {author}
		{\bibfnamefont {B.}~\bibnamefont {Rudek}}, \bibinfo {author} {\bibfnamefont
			{L.}~\bibnamefont {Foucar}}, \bibinfo {author} {\bibfnamefont {S.~H.}\
			\bibnamefont {Southworth}}, \bibinfo {author} {\bibfnamefont {C.~S.}\
			\bibnamefont {Lehmann}}, \bibinfo {author} {\bibfnamefont {B.}~\bibnamefont
			{Kraessig}}, \bibinfo {author} {\bibfnamefont {T.}~\bibnamefont {Marchenko}},
		\bibinfo {author} {\bibfnamefont {M.}~\bibnamefont {Simon}}, \bibinfo
		{author} {\bibfnamefont {K.}~\bibnamefont {Ueda}}, \bibinfo {author}
		{\bibfnamefont {K.~R.}\ \bibnamefont {Ferguson}}, \bibinfo {author}
		{\bibfnamefont {M.}~\bibnamefont {Bucher}}, \bibinfo {author} {\bibfnamefont
			{T.}~\bibnamefont {Gorkhover}}, \bibinfo {author} {\bibfnamefont
			{S.}~\bibnamefont {Carron}}, \bibinfo {author} {\bibfnamefont
			{R.}~\bibnamefont {Alonso-Mori}}, \bibinfo {author} {\bibfnamefont {J.~E.}\
			\bibnamefont {Koglin}}, \bibinfo {author} {\bibfnamefont {J.}~\bibnamefont
			{Correa}}, \bibinfo {author} {\bibfnamefont {G.~J.}\ \bibnamefont
			{Williams}}, \bibinfo {author} {\bibfnamefont {S.}~\bibnamefont {Boutet}},
		\bibinfo {author} {\bibfnamefont {L.}~\bibnamefont {Young}}, \bibinfo
		{author} {\bibfnamefont {C.}~\bibnamefont {Bostedt}}, \bibinfo {author}
		{\bibfnamefont {S.-K.}\ \bibnamefont {Son}}, \bibinfo {author} {\bibfnamefont
			{R.}~\bibnamefont {Santra}}, \ and\ \bibinfo {author} {\bibfnamefont
			{D.}~\bibnamefont {Rolles}},\ }\href
	{https://www.nature.com/articles/nature22373} {\bibfield  {journal} {\bibinfo
			{journal} {Nature}\ }\textbf {\bibinfo {volume} {546}},\ \bibinfo {pages}
		{129} (\bibinfo {year} {2017})}\BibitemShut {NoStop}%
	\bibitem [{\citenamefont {Fukuzawa}\ \emph {et~al.}(2013)\citenamefont
		{Fukuzawa}, \citenamefont {Son}, \citenamefont {Motomura}, \citenamefont
		{Mondal}, \citenamefont {Nagaya}, \citenamefont {Wada}, \citenamefont {Liu},
		\citenamefont {Feifel}, \citenamefont {Tachibana}, \citenamefont {Ito},
		\citenamefont {Kimura}, \citenamefont {Sakai}, \citenamefont {Matsunami},
		\citenamefont {Hayashita}, \citenamefont {Kajikawa}, \citenamefont
		{Johnsson}, \citenamefont {Siano}, \citenamefont {Kukk}, \citenamefont
		{Rudek}, \citenamefont {Erk}, \citenamefont {Foucar}, \citenamefont {Robert},
		\citenamefont {Miron}, \citenamefont {Tono}, \citenamefont {Inubushi},
		\citenamefont {Hatsui}, \citenamefont {Yabashi}, \citenamefont {Yao},
		\citenamefont {Santra},\ and\ \citenamefont {Ueda}}]{fukuzawa_deep_2013}%
	\BibitemOpen
	\bibfield  {author} {\bibinfo {author} {\bibfnamefont {H.}~\bibnamefont
			{Fukuzawa}}, \bibinfo {author} {\bibfnamefont {S.-K.}\ \bibnamefont {Son}},
		\bibinfo {author} {\bibfnamefont {K.}~\bibnamefont {Motomura}}, \bibinfo
		{author} {\bibfnamefont {S.}~\bibnamefont {Mondal}}, \bibinfo {author}
		{\bibfnamefont {K.}~\bibnamefont {Nagaya}}, \bibinfo {author} {\bibfnamefont
			{S.}~\bibnamefont {Wada}}, \bibinfo {author} {\bibfnamefont {X.-J.}\
			\bibnamefont {Liu}}, \bibinfo {author} {\bibfnamefont {R.}~\bibnamefont
			{Feifel}}, \bibinfo {author} {\bibfnamefont {T.}~\bibnamefont {Tachibana}},
		\bibinfo {author} {\bibfnamefont {Y.}~\bibnamefont {Ito}}, \bibinfo {author}
		{\bibfnamefont {M.}~\bibnamefont {Kimura}}, \bibinfo {author} {\bibfnamefont
			{T.}~\bibnamefont {Sakai}}, \bibinfo {author} {\bibfnamefont
			{K.}~\bibnamefont {Matsunami}}, \bibinfo {author} {\bibfnamefont
			{H.}~\bibnamefont {Hayashita}}, \bibinfo {author} {\bibfnamefont
			{J.}~\bibnamefont {Kajikawa}}, \bibinfo {author} {\bibfnamefont
			{P.}~\bibnamefont {Johnsson}}, \bibinfo {author} {\bibfnamefont
			{M.}~\bibnamefont {Siano}}, \bibinfo {author} {\bibfnamefont
			{E.}~\bibnamefont {Kukk}}, \bibinfo {author} {\bibfnamefont {B.}~\bibnamefont
			{Rudek}}, \bibinfo {author} {\bibfnamefont {B.}~\bibnamefont {Erk}}, \bibinfo
		{author} {\bibfnamefont {L.}~\bibnamefont {Foucar}}, \bibinfo {author}
		{\bibfnamefont {E.}~\bibnamefont {Robert}}, \bibinfo {author} {\bibfnamefont
			{C.}~\bibnamefont {Miron}}, \bibinfo {author} {\bibfnamefont
			{K.}~\bibnamefont {Tono}}, \bibinfo {author} {\bibfnamefont {Y.}~\bibnamefont
			{Inubushi}}, \bibinfo {author} {\bibfnamefont {T.}~\bibnamefont {Hatsui}},
		\bibinfo {author} {\bibfnamefont {M.}~\bibnamefont {Yabashi}}, \bibinfo
		{author} {\bibfnamefont {M.}~\bibnamefont {Yao}}, \bibinfo {author}
		{\bibfnamefont {R.}~\bibnamefont {Santra}}, \ and\ \bibinfo {author}
		{\bibfnamefont {K.}~\bibnamefont {Ueda}},\ }\href
	{https://link.aps.org/doi/10.1103/PhysRevLett.110.173005} {\bibfield
		{journal} {\bibinfo  {journal} {Physical Review Letters}\ }\textbf {\bibinfo
			{volume} {110}},\ \bibinfo {pages} {173005} (\bibinfo {year}
		{2013})}\BibitemShut {NoStop}%
	\bibitem [{\citenamefont {Erk}\ \emph {et~al.}(2013)\citenamefont {Erk},
		\citenamefont {Rolles}, \citenamefont {Foucar}, \citenamefont {Rudek},
		\citenamefont {Epp}, \citenamefont {Cryle}, \citenamefont {Bostedt},
		\citenamefont {Schorb}, \citenamefont {Bozek}, \citenamefont {Rouzee},
		\citenamefont {Hundertmark}, \citenamefont {Marchenko}, \citenamefont
		{Simon}, \citenamefont {Filsinger}, \citenamefont {Christensen},
		\citenamefont {De}, \citenamefont {Trippel}, \citenamefont {Küpper},
		\citenamefont {Stapelfeldt}, \citenamefont {Wada}, \citenamefont {Ueda},
		\citenamefont {Swiggers}, \citenamefont {Messerschmidt}, \citenamefont
		{Schröter}, \citenamefont {Moshammer}, \citenamefont {Schlichting},
		\citenamefont {Ullrich},\ and\ \citenamefont {Rudenko}}]{erk_ultrafast_2013}%
	\BibitemOpen
	\bibfield  {author} {\bibinfo {author} {\bibfnamefont {B.}~\bibnamefont
			{Erk}}, \bibinfo {author} {\bibfnamefont {D.}~\bibnamefont {Rolles}},
		\bibinfo {author} {\bibfnamefont {L.}~\bibnamefont {Foucar}}, \bibinfo
		{author} {\bibfnamefont {B.}~\bibnamefont {Rudek}}, \bibinfo {author}
		{\bibfnamefont {S.~W.}\ \bibnamefont {Epp}}, \bibinfo {author} {\bibfnamefont
			{M.}~\bibnamefont {Cryle}}, \bibinfo {author} {\bibfnamefont
			{C.}~\bibnamefont {Bostedt}}, \bibinfo {author} {\bibfnamefont
			{S.}~\bibnamefont {Schorb}}, \bibinfo {author} {\bibfnamefont
			{J.}~\bibnamefont {Bozek}}, \bibinfo {author} {\bibfnamefont
			{A.}~\bibnamefont {Rouzee}}, \bibinfo {author} {\bibfnamefont
			{A.}~\bibnamefont {Hundertmark}}, \bibinfo {author} {\bibfnamefont
			{T.}~\bibnamefont {Marchenko}}, \bibinfo {author} {\bibfnamefont
			{M.}~\bibnamefont {Simon}}, \bibinfo {author} {\bibfnamefont
			{F.}~\bibnamefont {Filsinger}}, \bibinfo {author} {\bibfnamefont
			{L.}~\bibnamefont {Christensen}}, \bibinfo {author} {\bibfnamefont
			{S.}~\bibnamefont {De}}, \bibinfo {author} {\bibfnamefont {S.}~\bibnamefont
			{Trippel}}, \bibinfo {author} {\bibfnamefont {J.}~\bibnamefont {Küpper}},
		\bibinfo {author} {\bibfnamefont {H.}~\bibnamefont {Stapelfeldt}}, \bibinfo
		{author} {\bibfnamefont {S.}~\bibnamefont {Wada}}, \bibinfo {author}
		{\bibfnamefont {K.}~\bibnamefont {Ueda}}, \bibinfo {author} {\bibfnamefont
			{M.}~\bibnamefont {Swiggers}}, \bibinfo {author} {\bibfnamefont
			{M.}~\bibnamefont {Messerschmidt}}, \bibinfo {author} {\bibfnamefont {C.~D.}\
			\bibnamefont {Schröter}}, \bibinfo {author} {\bibfnamefont {R.}~\bibnamefont
			{Moshammer}}, \bibinfo {author} {\bibfnamefont {I.}~\bibnamefont
			{Schlichting}}, \bibinfo {author} {\bibfnamefont {J.}~\bibnamefont
			{Ullrich}}, \ and\ \bibinfo {author} {\bibfnamefont {A.}~\bibnamefont
			{Rudenko}},\ }\href {https://link.aps.org/doi/10.1103/PhysRevLett.110.053003}
	{\bibfield  {journal} {\bibinfo  {journal} {Physical Review Letters}\
		}\textbf {\bibinfo {volume} {110}},\ \bibinfo {pages} {053003} (\bibinfo
		{year} {2013})}\BibitemShut {NoStop}%
	\bibitem [{\citenamefont {Erk}\ \emph {et~al.}(2014)\citenamefont {Erk},
		\citenamefont {Boll}, \citenamefont {Trippel}, \citenamefont {Anielski},
		\citenamefont {Foucar}, \citenamefont {Rudek}, \citenamefont {Epp},
		\citenamefont {Coffee}, \citenamefont {Carron}, \citenamefont {Schorb},
		\citenamefont {Ferguson}, \citenamefont {Swiggers}, \citenamefont {Bozek},
		\citenamefont {Simon}, \citenamefont {Marchenko}, \citenamefont {Kupper},
		\citenamefont {Schlichting}, \citenamefont {Ullrich}, \citenamefont
		{Bostedt}, \citenamefont {Rolles},\ and\ \citenamefont
		{Rudenko}}]{erk_imaging_2014}%
	\BibitemOpen
	\bibfield  {author} {\bibinfo {author} {\bibfnamefont {B.}~\bibnamefont
			{Erk}}, \bibinfo {author} {\bibfnamefont {R.}~\bibnamefont {Boll}}, \bibinfo
		{author} {\bibfnamefont {S.}~\bibnamefont {Trippel}}, \bibinfo {author}
		{\bibfnamefont {D.}~\bibnamefont {Anielski}}, \bibinfo {author}
		{\bibfnamefont {L.}~\bibnamefont {Foucar}}, \bibinfo {author} {\bibfnamefont
			{B.}~\bibnamefont {Rudek}}, \bibinfo {author} {\bibfnamefont {S.~W.}\
			\bibnamefont {Epp}}, \bibinfo {author} {\bibfnamefont {R.}~\bibnamefont
			{Coffee}}, \bibinfo {author} {\bibfnamefont {S.}~\bibnamefont {Carron}},
		\bibinfo {author} {\bibfnamefont {S.}~\bibnamefont {Schorb}}, \bibinfo
		{author} {\bibfnamefont {K.~R.}\ \bibnamefont {Ferguson}}, \bibinfo {author}
		{\bibfnamefont {M.}~\bibnamefont {Swiggers}}, \bibinfo {author}
		{\bibfnamefont {J.~D.}\ \bibnamefont {Bozek}}, \bibinfo {author}
		{\bibfnamefont {M.}~\bibnamefont {Simon}}, \bibinfo {author} {\bibfnamefont
			{T.}~\bibnamefont {Marchenko}}, \bibinfo {author} {\bibfnamefont
			{J.}~\bibnamefont {Kupper}}, \bibinfo {author} {\bibfnamefont
			{I.}~\bibnamefont {Schlichting}}, \bibinfo {author} {\bibfnamefont
			{J.}~\bibnamefont {Ullrich}}, \bibinfo {author} {\bibfnamefont
			{C.}~\bibnamefont {Bostedt}}, \bibinfo {author} {\bibfnamefont
			{D.}~\bibnamefont {Rolles}}, \ and\ \bibinfo {author} {\bibfnamefont
			{A.}~\bibnamefont {Rudenko}},\ }\href
	{http://www.sciencemag.org/cgi/doi/10.1126/science.1253607} {\bibfield
		{journal} {\bibinfo  {journal} {Science}\ }\textbf {\bibinfo {volume}
			{345}},\ \bibinfo {pages} {288} (\bibinfo {year} {2014})}\BibitemShut
	{NoStop}%
	\bibitem [{\citenamefont {Motomura}\ \emph {et~al.}(2015)\citenamefont
		{Motomura}, \citenamefont {Kukk}, \citenamefont {Fukuzawa}, \citenamefont
		{Wada}, \citenamefont {Nagaya}, \citenamefont {Ohmura}, \citenamefont
		{Mondal}, \citenamefont {Tachibana}, \citenamefont {Ito}, \citenamefont
		{Koga}, \citenamefont {Sakai}, \citenamefont {Matsunami}, \citenamefont
		{Rudenko}, \citenamefont {Nicolas}, \citenamefont {Liu}, \citenamefont
		{Miron}, \citenamefont {Zhang}, \citenamefont {Jiang}, \citenamefont {Chen},
		\citenamefont {Anand}, \citenamefont {Kim}, \citenamefont {Tono},
		\citenamefont {Yabashi}, \citenamefont {Yao},\ and\ \citenamefont
		{Ueda}}]{motomura_charge_2015}%
	\BibitemOpen
	\bibfield  {author} {\bibinfo {author} {\bibfnamefont {K.}~\bibnamefont
			{Motomura}}, \bibinfo {author} {\bibfnamefont {E.}~\bibnamefont {Kukk}},
		\bibinfo {author} {\bibfnamefont {H.}~\bibnamefont {Fukuzawa}}, \bibinfo
		{author} {\bibfnamefont {S.-I.}\ \bibnamefont {Wada}}, \bibinfo {author}
		{\bibfnamefont {K.}~\bibnamefont {Nagaya}}, \bibinfo {author} {\bibfnamefont
			{S.}~\bibnamefont {Ohmura}}, \bibinfo {author} {\bibfnamefont
			{S.}~\bibnamefont {Mondal}}, \bibinfo {author} {\bibfnamefont
			{T.}~\bibnamefont {Tachibana}}, \bibinfo {author} {\bibfnamefont
			{Y.}~\bibnamefont {Ito}}, \bibinfo {author} {\bibfnamefont {R.}~\bibnamefont
			{Koga}}, \bibinfo {author} {\bibfnamefont {T.}~\bibnamefont {Sakai}},
		\bibinfo {author} {\bibfnamefont {K.}~\bibnamefont {Matsunami}}, \bibinfo
		{author} {\bibfnamefont {A.}~\bibnamefont {Rudenko}}, \bibinfo {author}
		{\bibfnamefont {C.}~\bibnamefont {Nicolas}}, \bibinfo {author} {\bibfnamefont
			{X.-J.}\ \bibnamefont {Liu}}, \bibinfo {author} {\bibfnamefont
			{C.}~\bibnamefont {Miron}}, \bibinfo {author} {\bibfnamefont
			{Y.}~\bibnamefont {Zhang}}, \bibinfo {author} {\bibfnamefont
			{Y.}~\bibnamefont {Jiang}}, \bibinfo {author} {\bibfnamefont
			{J.}~\bibnamefont {Chen}}, \bibinfo {author} {\bibfnamefont {M.}~\bibnamefont
			{Anand}}, \bibinfo {author} {\bibfnamefont {D.~E.}\ \bibnamefont {Kim}},
		\bibinfo {author} {\bibfnamefont {K.}~\bibnamefont {Tono}}, \bibinfo {author}
		{\bibfnamefont {M.}~\bibnamefont {Yabashi}}, \bibinfo {author} {\bibfnamefont
			{M.}~\bibnamefont {Yao}}, \ and\ \bibinfo {author} {\bibfnamefont
			{K.}~\bibnamefont {Ueda}},\ }\href
	{https://doi.org/10.1021/acs.jpclett.5b01205} {\bibfield  {journal} {\bibinfo
			{journal} {The Journal of Physical Chemistry Letters}\ }\textbf {\bibinfo
			{volume} {6}},\ \bibinfo {pages} {2944} (\bibinfo {year} {2015})}\BibitemShut
	{NoStop}%
	\bibitem [{\citenamefont {Nagaya}\ \emph {et~al.}(2016)\citenamefont {Nagaya},
		\citenamefont {Motomura}, \citenamefont {Kukk}, \citenamefont {Fukuzawa},
		\citenamefont {Wada}, \citenamefont {Tachibana}, \citenamefont {Ito},
		\citenamefont {Mondal}, \citenamefont {Sakai}, \citenamefont {Matsunami},
		\citenamefont {Koga}, \citenamefont {Ohmura}, \citenamefont {Takahashi},
		\citenamefont {Kanno}, \citenamefont {Rudenko}, \citenamefont {Nicolas},
		\citenamefont {Liu}, \citenamefont {Zhang}, \citenamefont {Chen},
		\citenamefont {Anand}, \citenamefont {Jiang}, \citenamefont {Kim},
		\citenamefont {Tono}, \citenamefont {Yabashi}, \citenamefont {Kono},
		\citenamefont {Miron}, \citenamefont {Yao},\ and\ \citenamefont
		{Ueda}}]{nagaya_ultrafast_2016}%
	\BibitemOpen
	\bibfield  {author} {\bibinfo {author} {\bibfnamefont {K.}~\bibnamefont
			{Nagaya}}, \bibinfo {author} {\bibfnamefont {K.}~\bibnamefont {Motomura}},
		\bibinfo {author} {\bibfnamefont {E.}~\bibnamefont {Kukk}}, \bibinfo {author}
		{\bibfnamefont {H.}~\bibnamefont {Fukuzawa}}, \bibinfo {author}
		{\bibfnamefont {S.}~\bibnamefont {Wada}}, \bibinfo {author} {\bibfnamefont
			{T.}~\bibnamefont {Tachibana}}, \bibinfo {author} {\bibfnamefont
			{Y.}~\bibnamefont {Ito}}, \bibinfo {author} {\bibfnamefont {S.}~\bibnamefont
			{Mondal}}, \bibinfo {author} {\bibfnamefont {T.}~\bibnamefont {Sakai}},
		\bibinfo {author} {\bibfnamefont {K.}~\bibnamefont {Matsunami}}, \bibinfo
		{author} {\bibfnamefont {R.}~\bibnamefont {Koga}}, \bibinfo {author}
		{\bibfnamefont {S.}~\bibnamefont {Ohmura}}, \bibinfo {author} {\bibfnamefont
			{Y.}~\bibnamefont {Takahashi}}, \bibinfo {author} {\bibfnamefont
			{M.}~\bibnamefont {Kanno}}, \bibinfo {author} {\bibfnamefont
			{A.}~\bibnamefont {Rudenko}}, \bibinfo {author} {\bibfnamefont
			{C.}~\bibnamefont {Nicolas}}, \bibinfo {author} {\bibfnamefont {X.-J.}\
			\bibnamefont {Liu}}, \bibinfo {author} {\bibfnamefont {Y.}~\bibnamefont
			{Zhang}}, \bibinfo {author} {\bibfnamefont {J.}~\bibnamefont {Chen}},
		\bibinfo {author} {\bibfnamefont {M.}~\bibnamefont {Anand}}, \bibinfo
		{author} {\bibfnamefont {Y.~H.}\ \bibnamefont {Jiang}}, \bibinfo {author}
		{\bibfnamefont {D.-E.}\ \bibnamefont {Kim}}, \bibinfo {author} {\bibfnamefont
			{K.}~\bibnamefont {Tono}}, \bibinfo {author} {\bibfnamefont {M.}~\bibnamefont
			{Yabashi}}, \bibinfo {author} {\bibfnamefont {H.}~\bibnamefont {Kono}},
		\bibinfo {author} {\bibfnamefont {C.}~\bibnamefont {Miron}}, \bibinfo
		{author} {\bibfnamefont {M.}~\bibnamefont {Yao}}, \ and\ \bibinfo {author}
		{\bibfnamefont {K.}~\bibnamefont {Ueda}},\ }\href
	{https://link.aps.org/doi/10.1103/PhysRevX.6.021035} {\bibfield  {journal}
		{\bibinfo  {journal} {Physical Review X}\ }\textbf {\bibinfo {volume} {6}},\
		\bibinfo {pages} {021035} (\bibinfo {year} {2016})}\BibitemShut {NoStop}%
	\bibitem [{\citenamefont {Neutze}\ \emph {et~al.}(2000)\citenamefont {Neutze},
		\citenamefont {Wouts}, \citenamefont {Spoel}, \citenamefont {Weckert},\ and\
		\citenamefont {Hajdu}}]{neutze_potential_2000}%
	\BibitemOpen
	\bibfield  {author} {\bibinfo {author} {\bibfnamefont {R.}~\bibnamefont
			{Neutze}}, \bibinfo {author} {\bibfnamefont {R.}~\bibnamefont {Wouts}},
		\bibinfo {author} {\bibfnamefont {D.~v.~d.}\ \bibnamefont {Spoel}}, \bibinfo
		{author} {\bibfnamefont {E.}~\bibnamefont {Weckert}}, \ and\ \bibinfo
		{author} {\bibfnamefont {J.}~\bibnamefont {Hajdu}},\ }\href
	{https://www.nature.com/articles/35021099} {\bibfield  {journal} {\bibinfo
			{journal} {Nature}\ }\textbf {\bibinfo {volume} {406}},\ \bibinfo {pages}
		{752} (\bibinfo {year} {2000})}\BibitemShut {NoStop}%
	\bibitem [{\citenamefont {Barty}\ \emph {et~al.}(2008)\citenamefont {Barty},
		\citenamefont {Boutet}, \citenamefont {Bogan}, \citenamefont {Hau-Riege},
		\citenamefont {Marchesini}, \citenamefont {Sokolowski-Tinten}, \citenamefont
		{Stojanovic}, \citenamefont {Tobey}, \citenamefont {Ehrke}, \citenamefont
		{Cavalleri}, \citenamefont {Düsterer}, \citenamefont {Frank}, \citenamefont
		{Bajt}, \citenamefont {Woods}, \citenamefont {Seibert}, \citenamefont
		{Hajdu}, \citenamefont {Treusch},\ and\ \citenamefont
		{Chapman}}]{barty_ultrafast_2008}%
	\BibitemOpen
	\bibfield  {author} {\bibinfo {author} {\bibfnamefont {A.}~\bibnamefont
			{Barty}}, \bibinfo {author} {\bibfnamefont {S.}~\bibnamefont {Boutet}},
		\bibinfo {author} {\bibfnamefont {M.~J.}\ \bibnamefont {Bogan}}, \bibinfo
		{author} {\bibfnamefont {S.}~\bibnamefont {Hau-Riege}}, \bibinfo {author}
		{\bibfnamefont {S.}~\bibnamefont {Marchesini}}, \bibinfo {author}
		{\bibfnamefont {K.}~\bibnamefont {Sokolowski-Tinten}}, \bibinfo {author}
		{\bibfnamefont {N.}~\bibnamefont {Stojanovic}}, \bibinfo {author}
		{\bibfnamefont {R.}~\bibnamefont {Tobey}}, \bibinfo {author} {\bibfnamefont
			{H.}~\bibnamefont {Ehrke}}, \bibinfo {author} {\bibfnamefont
			{A.}~\bibnamefont {Cavalleri}}, \bibinfo {author} {\bibfnamefont
			{S.}~\bibnamefont {Düsterer}}, \bibinfo {author} {\bibfnamefont
			{M.}~\bibnamefont {Frank}}, \bibinfo {author} {\bibfnamefont
			{S.}~\bibnamefont {Bajt}}, \bibinfo {author} {\bibfnamefont {B.~W.}\
			\bibnamefont {Woods}}, \bibinfo {author} {\bibfnamefont {M.~M.}\ \bibnamefont
			{Seibert}}, \bibinfo {author} {\bibfnamefont {J.}~\bibnamefont {Hajdu}},
		\bibinfo {author} {\bibfnamefont {R.}~\bibnamefont {Treusch}}, \ and\
		\bibinfo {author} {\bibfnamefont {H.~N.}\ \bibnamefont {Chapman}},\ }\href
	{https://www.nature.com/articles/nphoton.2008.128} {\bibfield  {journal}
		{\bibinfo  {journal} {Nature Photonics}\ }\textbf {\bibinfo {volume} {2}},\
		\bibinfo {pages} {415} (\bibinfo {year} {2008})}\BibitemShut {NoStop}%
	\bibitem [{\citenamefont {Nass}\ \emph {et~al.}(2015)\citenamefont {Nass},
		\citenamefont {Foucar}, \citenamefont {Barends}, \citenamefont {Hartmann},
		\citenamefont {Botha}, \citenamefont {Shoeman}, \citenamefont {Doak},
		\citenamefont {Alonso-Mori}, \citenamefont {Aquila}, \citenamefont {Bajt},
		\citenamefont {Barty}, \citenamefont {Bean}, \citenamefont {Beyerlein},
		\citenamefont {Bublitz}, \citenamefont {Drachmann}, \citenamefont
		{Gregersen}, \citenamefont {Jönsson}, \citenamefont {Kabsch}, \citenamefont
		{Kassemeyer}, \citenamefont {Koglin}, \citenamefont {Krumrey}, \citenamefont
		{Mattle}, \citenamefont {Messerschmidt}, \citenamefont {Nissen},
		\citenamefont {Reinhard}, \citenamefont {Sitsel}, \citenamefont {Sokaras},
		\citenamefont {Williams}, \citenamefont {Hau-Riege}, \citenamefont
		{Timneanu}, \citenamefont {Caleman}, \citenamefont {Chapman}, \citenamefont
		{Boutet},\ and\ \citenamefont {Schlichting}}]{nass_indications_2015}%
	\BibitemOpen
	\bibfield  {author} {\bibinfo {author} {\bibfnamefont {K.}~\bibnamefont
			{Nass}}, \bibinfo {author} {\bibfnamefont {L.}~\bibnamefont {Foucar}},
		\bibinfo {author} {\bibfnamefont {T.~R.~M.}\ \bibnamefont {Barends}},
		\bibinfo {author} {\bibfnamefont {E.}~\bibnamefont {Hartmann}}, \bibinfo
		{author} {\bibfnamefont {S.}~\bibnamefont {Botha}}, \bibinfo {author}
		{\bibfnamefont {R.~L.}\ \bibnamefont {Shoeman}}, \bibinfo {author}
		{\bibfnamefont {R.~B.}\ \bibnamefont {Doak}}, \bibinfo {author}
		{\bibfnamefont {R.}~\bibnamefont {Alonso-Mori}}, \bibinfo {author}
		{\bibfnamefont {A.}~\bibnamefont {Aquila}}, \bibinfo {author} {\bibfnamefont
			{S.}~\bibnamefont {Bajt}}, \bibinfo {author} {\bibfnamefont {A.}~\bibnamefont
			{Barty}}, \bibinfo {author} {\bibfnamefont {R.}~\bibnamefont {Bean}},
		\bibinfo {author} {\bibfnamefont {K.~R.}\ \bibnamefont {Beyerlein}}, \bibinfo
		{author} {\bibfnamefont {M.}~\bibnamefont {Bublitz}}, \bibinfo {author}
		{\bibfnamefont {N.}~\bibnamefont {Drachmann}}, \bibinfo {author}
		{\bibfnamefont {J.}~\bibnamefont {Gregersen}}, \bibinfo {author}
		{\bibfnamefont {H.~O.}\ \bibnamefont {Jönsson}}, \bibinfo {author}
		{\bibfnamefont {W.}~\bibnamefont {Kabsch}}, \bibinfo {author} {\bibfnamefont
			{S.}~\bibnamefont {Kassemeyer}}, \bibinfo {author} {\bibfnamefont {J.~E.}\
			\bibnamefont {Koglin}}, \bibinfo {author} {\bibfnamefont {M.}~\bibnamefont
			{Krumrey}}, \bibinfo {author} {\bibfnamefont {D.}~\bibnamefont {Mattle}},
		\bibinfo {author} {\bibfnamefont {M.}~\bibnamefont {Messerschmidt}}, \bibinfo
		{author} {\bibfnamefont {P.}~\bibnamefont {Nissen}}, \bibinfo {author}
		{\bibfnamefont {L.}~\bibnamefont {Reinhard}}, \bibinfo {author}
		{\bibfnamefont {O.}~\bibnamefont {Sitsel}}, \bibinfo {author} {\bibfnamefont
			{D.}~\bibnamefont {Sokaras}}, \bibinfo {author} {\bibfnamefont {G.~J.}\
			\bibnamefont {Williams}}, \bibinfo {author} {\bibfnamefont {S.}~\bibnamefont
			{Hau-Riege}}, \bibinfo {author} {\bibfnamefont {N.}~\bibnamefont {Timneanu}},
		\bibinfo {author} {\bibfnamefont {C.}~\bibnamefont {Caleman}}, \bibinfo
		{author} {\bibfnamefont {H.~N.}\ \bibnamefont {Chapman}}, \bibinfo {author}
		{\bibfnamefont {S.}~\bibnamefont {Boutet}}, \ and\ \bibinfo {author}
		{\bibfnamefont {I.}~\bibnamefont {Schlichting}},\ }\href@noop {} {\bibfield
		{journal} {\bibinfo  {journal} {Journal of Synchrotron Radiation}\ }\textbf
		{\bibinfo {volume} {22}},\ \bibinfo {pages} {225} (\bibinfo {year}
		{2015})}\BibitemShut {NoStop}%
	\bibitem [{\citenamefont {Galli}\ \emph {et~al.}(2015)\citenamefont {Galli},
		\citenamefont {Son}, \citenamefont {Klinge}, \citenamefont {Bajt},
		\citenamefont {Barty}, \citenamefont {Bean}, \citenamefont {Betzel},
		\citenamefont {Beyerlein}, \citenamefont {Caleman}, \citenamefont {Doak},
		\citenamefont {Duszenko}, \citenamefont {Fleckenstein}, \citenamefont {Gati},
		\citenamefont {Hunt}, \citenamefont {Kirian}, \citenamefont {Liang},
		\citenamefont {Nanao}, \citenamefont {Nass}, \citenamefont {Oberthür},
		\citenamefont {Redecke}, \citenamefont {Shoeman}, \citenamefont {Stellato},
		\citenamefont {Yoon}, \citenamefont {White}, \citenamefont {Yefanov},
		\citenamefont {Spence},\ and\ \citenamefont
		{Chapman}}]{galli_electronic_2015}%
	\BibitemOpen
	\bibfield  {author} {\bibinfo {author} {\bibfnamefont {L.}~\bibnamefont
			{Galli}}, \bibinfo {author} {\bibfnamefont {S.-K.}\ \bibnamefont {Son}},
		\bibinfo {author} {\bibfnamefont {M.}~\bibnamefont {Klinge}}, \bibinfo
		{author} {\bibfnamefont {S.}~\bibnamefont {Bajt}}, \bibinfo {author}
		{\bibfnamefont {A.}~\bibnamefont {Barty}}, \bibinfo {author} {\bibfnamefont
			{R.}~\bibnamefont {Bean}}, \bibinfo {author} {\bibfnamefont {C.}~\bibnamefont
			{Betzel}}, \bibinfo {author} {\bibfnamefont {K.~R.}\ \bibnamefont
			{Beyerlein}}, \bibinfo {author} {\bibfnamefont {C.}~\bibnamefont {Caleman}},
		\bibinfo {author} {\bibfnamefont {R.~B.}\ \bibnamefont {Doak}}, \bibinfo
		{author} {\bibfnamefont {M.}~\bibnamefont {Duszenko}}, \bibinfo {author}
		{\bibfnamefont {H.}~\bibnamefont {Fleckenstein}}, \bibinfo {author}
		{\bibfnamefont {C.}~\bibnamefont {Gati}}, \bibinfo {author} {\bibfnamefont
			{B.}~\bibnamefont {Hunt}}, \bibinfo {author} {\bibfnamefont {R.~A.}\
			\bibnamefont {Kirian}}, \bibinfo {author} {\bibfnamefont {M.}~\bibnamefont
			{Liang}}, \bibinfo {author} {\bibfnamefont {M.~H.}\ \bibnamefont {Nanao}},
		\bibinfo {author} {\bibfnamefont {K.}~\bibnamefont {Nass}}, \bibinfo {author}
		{\bibfnamefont {D.}~\bibnamefont {Oberthür}}, \bibinfo {author}
		{\bibfnamefont {L.}~\bibnamefont {Redecke}}, \bibinfo {author} {\bibfnamefont
			{R.}~\bibnamefont {Shoeman}}, \bibinfo {author} {\bibfnamefont
			{F.}~\bibnamefont {Stellato}}, \bibinfo {author} {\bibfnamefont {C.~H.}\
			\bibnamefont {Yoon}}, \bibinfo {author} {\bibfnamefont {T.~A.}\ \bibnamefont
			{White}}, \bibinfo {author} {\bibfnamefont {O.}~\bibnamefont {Yefanov}},
		\bibinfo {author} {\bibfnamefont {J.}~\bibnamefont {Spence}}, \ and\ \bibinfo
		{author} {\bibfnamefont {H.~N.}\ \bibnamefont {Chapman}},\ }\href
	{https://www.ncbi.nlm.nih.gov/pmc/articles/PMC4711609/} {\bibfield  {journal}
		{\bibinfo  {journal} {Structural Dynamics}\ }\textbf {\bibinfo {volume} {2}}
		(\bibinfo {year} {2015})}\BibitemShut {NoStop}%
	\bibitem [{\citenamefont {Rohringer}\ and\ \citenamefont
		{Santra}(2007)}]{rohringer_x-ray_2007}%
	\BibitemOpen
	\bibfield  {author} {\bibinfo {author} {\bibfnamefont {N.}~\bibnamefont
			{Rohringer}}\ and\ \bibinfo {author} {\bibfnamefont {R.}~\bibnamefont
			{Santra}},\ }\href {https://link.aps.org/doi/10.1103/PhysRevA.76.033416}
	{\bibfield  {journal} {\bibinfo  {journal} {Physical Review A}\ }\textbf
		{\bibinfo {volume} {76}},\ \bibinfo {pages} {033416} (\bibinfo {year}
		{2007})}\BibitemShut {NoStop}%
	\bibitem [{\citenamefont {Son}\ and\ \citenamefont
		{Santra}(2012)}]{son_monte_2012}%
	\BibitemOpen
	\bibfield  {author} {\bibinfo {author} {\bibfnamefont {S.-K.}\ \bibnamefont
			{Son}}\ and\ \bibinfo {author} {\bibfnamefont {R.}~\bibnamefont {Santra}},\
	}\href {https://link.aps.org/doi/10.1103/PhysRevA.85.063415} {\bibfield
	{journal} {\bibinfo  {journal} {Physical Review A}\ }\textbf {\bibinfo
		{volume} {85}},\ \bibinfo {pages} {063415} (\bibinfo {year}
	{2012})}\BibitemShut {NoStop}%
\bibitem [{\citenamefont {Ho}\ \emph {et~al.}(2014)\citenamefont {Ho},
	\citenamefont {Bostedt}, \citenamefont {Schorb},\ and\ \citenamefont
	{Young}}]{ho_theoretical_2014}%
\BibitemOpen
\bibfield  {author} {\bibinfo {author} {\bibfnamefont {P.~J.}\ \bibnamefont
		{Ho}}, \bibinfo {author} {\bibfnamefont {C.}~\bibnamefont {Bostedt}},
	\bibinfo {author} {\bibfnamefont {S.}~\bibnamefont {Schorb}}, \ and\ \bibinfo
	{author} {\bibfnamefont {L.}~\bibnamefont {Young}},\ }\href
{https://link.aps.org/doi/10.1103/PhysRevLett.113.253001} {\bibfield
	{journal} {\bibinfo  {journal} {Physical Review Letters}\ }\textbf {\bibinfo
		{volume} {113}},\ \bibinfo {pages} {253001} (\bibinfo {year}
	{2014})}\BibitemShut {NoStop}%
\bibitem [{\citenamefont {Hao}\ \emph {et~al.}(2015)\citenamefont {Hao},
	\citenamefont {Inhester}, \citenamefont {Hanasaki}, \citenamefont {Son},\
	and\ \citenamefont {Santra}}]{hao_efficient_2015}%
\BibitemOpen
\bibfield  {author} {\bibinfo {author} {\bibfnamefont {Y.}~\bibnamefont
		{Hao}}, \bibinfo {author} {\bibfnamefont {L.}~\bibnamefont {Inhester}},
	\bibinfo {author} {\bibfnamefont {K.}~\bibnamefont {Hanasaki}}, \bibinfo
	{author} {\bibfnamefont {S.-K.}\ \bibnamefont {Son}}, \ and\ \bibinfo
	{author} {\bibfnamefont {R.}~\bibnamefont {Santra}},\ }\href
{https://aca.scitation.org/doi/abs/10.1063/1.4919794} {\bibfield  {journal}
	{\bibinfo  {journal} {Structural Dynamics}\ }\textbf {\bibinfo {volume}
		{2}},\ \bibinfo {pages} {041707} (\bibinfo {year} {2015})}\BibitemShut
{NoStop}%
\bibitem [{\citenamefont {Inhester}\ \emph {et~al.}(2016)\citenamefont
	{Inhester}, \citenamefont {Hanasaki}, \citenamefont {Hao}, \citenamefont
	{Son},\ and\ \citenamefont {Santra}}]{inhester_x-ray_2016}%
\BibitemOpen
\bibfield  {author} {\bibinfo {author} {\bibfnamefont {L.}~\bibnamefont
		{Inhester}}, \bibinfo {author} {\bibfnamefont {K.}~\bibnamefont {Hanasaki}},
	\bibinfo {author} {\bibfnamefont {Y.}~\bibnamefont {Hao}}, \bibinfo {author}
	{\bibfnamefont {S.-K.}\ \bibnamefont {Son}}, \ and\ \bibinfo {author}
	{\bibfnamefont {R.}~\bibnamefont {Santra}},\ }\href
{https://link.aps.org/doi/10.1103/PhysRevA.94.023422} {\bibfield  {journal}
	{\bibinfo  {journal} {Physical Review A}\ }\textbf {\bibinfo {volume} {94}},\
	\bibinfo {pages} {023422} (\bibinfo {year} {2016})}\BibitemShut {NoStop}%
\bibitem [{\citenamefont {Rudek}\ \emph {et~al.}(2018)\citenamefont {Rudek},
	\citenamefont {Toyota}, \citenamefont {Foucar}, \citenamefont {Erk},
	\citenamefont {Boll}, \citenamefont {Bomme}, \citenamefont {Correa},
	\citenamefont {Carron}, \citenamefont {Boutet}, \citenamefont {Williams},
	\citenamefont {Ferguson}, \citenamefont {Alonso-Mori}, \citenamefont
	{Koglin}, \citenamefont {Gorkhover}, \citenamefont {Bucher}, \citenamefont
	{Lehmann}, \citenamefont {Krässig}, \citenamefont {Southworth},
	\citenamefont {Young}, \citenamefont {Bostedt}, \citenamefont {Ueda},
	\citenamefont {Marchenko}, \citenamefont {Simon}, \citenamefont {Jurek},
	\citenamefont {Santra}, \citenamefont {Rudenko}, \citenamefont {Son},\ and\
	\citenamefont {Rolles}}]{rudek_relativistic_2018}%
\BibitemOpen
\bibfield  {author} {\bibinfo {author} {\bibfnamefont {B.}~\bibnamefont
		{Rudek}}, \bibinfo {author} {\bibfnamefont {K.}~\bibnamefont {Toyota}},
	\bibinfo {author} {\bibfnamefont {L.}~\bibnamefont {Foucar}}, \bibinfo
	{author} {\bibfnamefont {B.}~\bibnamefont {Erk}}, \bibinfo {author}
	{\bibfnamefont {R.}~\bibnamefont {Boll}}, \bibinfo {author} {\bibfnamefont
		{C.}~\bibnamefont {Bomme}}, \bibinfo {author} {\bibfnamefont
		{J.}~\bibnamefont {Correa}}, \bibinfo {author} {\bibfnamefont
		{S.}~\bibnamefont {Carron}}, \bibinfo {author} {\bibfnamefont
		{S.}~\bibnamefont {Boutet}}, \bibinfo {author} {\bibfnamefont {G.~J.}\
		\bibnamefont {Williams}}, \bibinfo {author} {\bibfnamefont {K.~R.}\
		\bibnamefont {Ferguson}}, \bibinfo {author} {\bibfnamefont {R.}~\bibnamefont
		{Alonso-Mori}}, \bibinfo {author} {\bibfnamefont {J.~E.}\ \bibnamefont
		{Koglin}}, \bibinfo {author} {\bibfnamefont {T.}~\bibnamefont {Gorkhover}},
	\bibinfo {author} {\bibfnamefont {M.}~\bibnamefont {Bucher}}, \bibinfo
	{author} {\bibfnamefont {C.~S.}\ \bibnamefont {Lehmann}}, \bibinfo {author}
	{\bibfnamefont {B.}~\bibnamefont {Krässig}}, \bibinfo {author}
	{\bibfnamefont {S.~H.}\ \bibnamefont {Southworth}}, \bibinfo {author}
	{\bibfnamefont {L.}~\bibnamefont {Young}}, \bibinfo {author} {\bibfnamefont
		{C.}~\bibnamefont {Bostedt}}, \bibinfo {author} {\bibfnamefont
		{K.}~\bibnamefont {Ueda}}, \bibinfo {author} {\bibfnamefont {T.}~\bibnamefont
		{Marchenko}}, \bibinfo {author} {\bibfnamefont {M.}~\bibnamefont {Simon}},
	\bibinfo {author} {\bibfnamefont {Z.}~\bibnamefont {Jurek}}, \bibinfo
	{author} {\bibfnamefont {R.}~\bibnamefont {Santra}}, \bibinfo {author}
	{\bibfnamefont {A.}~\bibnamefont {Rudenko}}, \bibinfo {author} {\bibfnamefont
		{S.-K.}\ \bibnamefont {Son}}, \ and\ \bibinfo {author} {\bibfnamefont
		{D.}~\bibnamefont {Rolles}},\ }\href
{https://www.nature.com/articles/s41467-018-06745-6} {\bibfield  {journal}
	{\bibinfo  {journal} {Nature Communications}\ }\textbf {\bibinfo {volume}
		{9}},\ \bibinfo {pages} {4200} (\bibinfo {year} {2018})}\BibitemShut
{NoStop}%
\bibitem [{li_(view{\natexlab{a}})}]{li_pulse_inrev}%
\BibitemOpen
\href@noop {} {\bibfield  {journal} {\bibinfo  {journal} {X. Li, L. Inhester,
			S.J. Robatjazi, B. Erk, R. Boll, K. Hanasaki, K. Toyota, Y. Hao, C. Bomme, B.
			Rudek, L. Foucar, S. H. Southworth, C. S. Lehmann, B. Kraessig, T. Marchenko,
			M. Simon, K. Ueda, K. R. Ferguson, M. Bucher, T. Gorkhover, S. Carron, R.
			Alonso-Mori, J.E. Koglin, J. Correa, G. J. Williams, S. Boutet, L. Young, C.
			Bostedt, S.-K. Son, R. Santra, D. Rolles and A. Rudenko, Pulse energy and
			pulse duration effects in the ionization and fragmentation of polyatomic
			molecules by ultra-intense hard x-rays}\ } (\bibinfo {year} {in
		review}{\natexlab{a}})}\BibitemShut {NoStop}%
\bibitem [{li_(view{\natexlab{b}})}]{li_coinc_inrev}%
\BibitemOpen
\href@noop {} {\bibfield  {journal} {\bibinfo  {journal} {X. Li, L. Inhester,
			T. Osipov, R. Boll, R. Coffee, J. Cryan, A. Gatton, T. Gorkhover, G. Hartman,
			M. Ilchen, A. Knie, M.-F. Lin, M. P. Minitti, C. Weninger, T. J. A. Wolf,
			S.-K. Son, R. Santra, D. Rolles, A. Rudenko and P. Walter, Electron-ion
			coincidence measurements of molecular dynamics with intense x-ray pulses}\ }
	(\bibinfo {year} {in review}{\natexlab{b}})}\BibitemShut {NoStop}%
\bibitem [{\citenamefont {Guichard}\ \emph {et~al.}(2013)\citenamefont
	{Guichard}, \citenamefont {Richter}, \citenamefont {Rost}, \citenamefont
	{Saalmann}, \citenamefont {Sorokin},\ and\ \citenamefont
	{Tiedtke}}]{guichard_multiple_2013}%
\BibitemOpen
\bibfield  {author} {\bibinfo {author} {\bibfnamefont {R.}~\bibnamefont
		{Guichard}}, \bibinfo {author} {\bibfnamefont {M.}~\bibnamefont {Richter}},
	\bibinfo {author} {\bibfnamefont {J.-M.}\ \bibnamefont {Rost}}, \bibinfo
	{author} {\bibfnamefont {U.}~\bibnamefont {Saalmann}}, \bibinfo {author}
	{\bibfnamefont {A.~A.}\ \bibnamefont {Sorokin}}, \ and\ \bibinfo {author}
	{\bibfnamefont {K.}~\bibnamefont {Tiedtke}},\ }\href
{https://doi.org/10.1088%2F0953-4075%2F46%2F16%2F164025} {\bibfield
	{journal} {\bibinfo  {journal} {Journal of Physics B: Atomic, Molecular and
			Optical Physics}\ }\textbf {\bibinfo {volume} {46}},\ \bibinfo {pages}
	{164025} (\bibinfo {year} {2013})},\ \bibinfo {note} {publisher: IOP
	Publishing}\BibitemShut {NoStop}%
\bibitem [{\citenamefont {Li}(2019)}]{li_molecular_2019}%
\BibitemOpen
\bibfield  {author} {\bibinfo {author} {\bibfnamefont {X.}~\bibnamefont
		{Li}},\ }\emph {\bibinfo {title} {Section 6.3, Molecular response to
		ultra-intense x-rays studied with ion and electron momentum imaging}},\ \href
{https://krex.k-state.edu/dspace/bitstream/handle/2097/40257/XiangLi2019.pdf?sequence=3&isAllowed=y}
{Ph.D. thesis},\ \bibinfo  {school} {Kansas State University} (\bibinfo
{year} {2019})\BibitemShut {NoStop}%
\bibitem [{\citenamefont {Cowan}()}]{cowan_code}%
\BibitemOpen
\bibfield  {author} {\bibinfo {author} {\bibfnamefont {R.~D.}\ \bibnamefont
		{Cowan}},\ }\href
{https://www.tcd.ie/Physics/people/Cormac.McGuinness/Cowan/} {\emph {\bibinfo
		{title} {Atomic Structure Codes}}}\BibitemShut {NoStop}%
\bibitem [{\citenamefont {Cowan}(1981)}]{cowan_book_1981}%
\BibitemOpen
\bibfield  {author} {\bibinfo {author} {\bibfnamefont {R.~D.}\ \bibnamefont
		{Cowan}},\ }\href@noop {} {\emph {\bibinfo {title} {The Theory of Atomic
			Structure and Spectra}}}\ (\bibinfo  {publisher} {University of California
	Press, Berkeley, CA},\ \bibinfo {year} {1981})\BibitemShut {NoStop}%
\bibitem [{\citenamefont {Berger}\ \emph {et~al.}()\citenamefont {Berger} \emph
	{et~al.}}]{berger_xcom_1998}%
\BibitemOpen
\bibfield  {author} {\bibinfo {author} {\bibfnamefont {M.}~\bibnamefont
		{Berger}} \emph {et~al.},\ }\href
{https://www.nist.gov/pml/xcom-photon-cross-sections-database} {\emph
	{\bibinfo {title} {{XCOM}: {Photon} {Cross} {Sections} {Database},
			1998}}}\BibitemShut {NoStop}%
\bibitem [{li_(tion)}]{li_coulomb_tosub}%
\BibitemOpen
\href@noop {} {\bibfield  {journal} {\bibinfo  {journal} {X. Li, A. Rudenko,
			K. Fehre, G. Kastirke, M. S. Sch{\"o}ffler, M. M. Abdullah, N. Anders, T. M.
			Baumann, A. Czasch, S. Eckart, B. Erk, A. De Fanis, R. D{\"o}rner, L. Foucar,
			S. Grundmann, P. Grychtol, A. Hartung, M. Hofmann, M. Ilchen,, C. Janke, M.
			Kircher, K. Kubicek, M. Kunitski, T. Mazza, S. Meister, N. Melzer, J.
			Montano, V. Music, G. Nalin, Y. Ovcharenko,C. Passow, A. Pier, N. Rennhack,
			J. Rist, D. E. Rivas, I. Schlichting, L. Ph. H. Schmidt, Ph. Schmidt, J.
			Siebert, N. Strenger, D. Trabert, F. Trinter, I. Vela-Perez, R. Wagner, P.
			Walter, M. Weller, P. Ziolkowski, D. Rolles, M. Meyer, T. Jahnke and R. Boll,
			Coulomb explosion imaging of small polyatomic molecules with ultrashort x-ray
			pulses}\ } (\bibinfo {year} {in preparation})}\BibitemShut {NoStop}%
\end{thebibliography}
%

\end{document}